\documentclass[preprint,12pt]{elsarticle}

\usepackage{times,amsmath,epsfig}
\usepackage{tikz}
\usepackage{booktabs}
\usepackage{tablefootnote}
\usepackage{fancyhdr}
\usepackage{amsmath}
\usepackage{amsfonts}
\usepackage{amssymb}
\usepackage{array}
\usepackage{graphicx}
\usepackage{url}
\usepackage{accessibility}
\usepackage{subcaption}
\usepackage{bm}
\usepackage{breqn}
\usepackage{xcolor}
\usepackage{soul}
\usepackage{amssymb}
\usepackage[breaklinks=true]{hyperref}
\usepackage{breakcites}
\usepackage{hyperref}
\usepackage{multirow}
\usepackage{bm}
\usepackage{everypage}

\usepackage{natbib}
\setcitestyle{square}
\setcitestyle{numbers}

\usepackage[acronyms, nonumberlist, nopostdot, nomain, nogroupskip]{glossaries}

\newacronym{ez}{EZ}{Epileptogenic Zone}
\newacronym{pz}{PZ}{Propagation Zone}
\newacronym{dez}{D-EZ}{Detectable Epileptogenic Zone}
\newacronym{dpz}{D-PZ}{Detectable Propagation Zone}
\newacronym{seeg}{SEEG}{Stereoelectroencephalography}
\newacronym{ei}{EI}{Epileptogenicity Index}
\newacronym{em}{EM}{Epileptogenic Map}
\newacronym{ccep}{CCEP}{Cortico-Cortical Evoked Potential}
\newacronym{plv}{PLV}{Phase Locking Value}
\newacronym{pli}{PLI}{Phase Lag Index}
\newacronym{gc}{GC}{Granger Casuality}
\newacronym{pte}{PTE}{Phase Transfer Entropy}
\newacronym{di}{DI}{Desynchronization Index}
\newacronym{si}{SI}{Synchronization Index}
\newacronym{ft}{FT}{Fourier Transform}
\newacronym{ht}{HT}{Hilbert Transform}
\newacronym{ewma}{EWMA}{Exponentially Weighted Moving Average}
\newacronym{cusum}{CUSUM}{CUmulative SUM}
\newacronym{wm}{WM}{White Matter}
\newacronym{ml}{ML}{Machine Learning}
\newacronym{svm}{SVM}{Support Vector Machine}
\newacronym{auc}{ROC-AUC}{Area Under the ROC Curve}
\newacronym{pr-auc}{PR-AUC}{Area Under the Precision-Recall Curve}
\newacronym{roc}{ROC}{Receiver Operating Characteristic}
\newacronym{rftc}{RFTC}{Radio Frequency Thermocoagulation}

\DeclareMathOperator*{\argmax}{arg\,max}

\journal{Biomedical Signal Processing and Control}

\begin{document}


\begin{frontmatter}

\title{\vspace{-3.in}Desynchronization Index: A New Connectivity Approach for Exploring Epileptogenic Networks}

\author[unipd]{Federico Mason}

\author[irccs]{Lorenzo Ferri\corref{cor1}}

\author[irccs]{Lidia Di Vito}

\author[unibo]{Lara Alvisi}

\author[unibo]{Luca Zanuttini}

\author[irccs]{Matteo Martinoni}

\author[niguarda]{Roberto Mai}

\author[niguarda, unipa]{Francesco Cardinale}

\author[unibo]{Paolo Tinuper}

\author[irccs]{Roberto Michelucci}

\author[irccs]{Elena Pasini\corref{cor2}}

\author[irccs,unibo]{Francesca Bisulli\corref{cor2}}

\affiliation[unipd]{organization={Department of Information Engineering, University of Padova}, city={Padova}, country={Italy}}

\affiliation[irccs]{organization={IRCCS Institute of Neurological Sciences of Bologna, European Reference Network for Rare and Complex Epilepsies (EpiCARE)}, city={Bologna}, country={Italy}}

\affiliation[unibo]{organization={Department of Neuromotor and Biomedical Sciences, University of Bologna}, city={Bologna}, country={Italy}}

\affiliation[niguarda]{organization={Claudio Munari Epilepsy and Parkinson Surgery Centre, Niguarda Hospital}, city={Milan}, country={Italy}}

\affiliation[unipa]{organization={Department of Medicine and Surgery, University of Parma}, city={Parma}, country={Italy}}

\cortext[cor1]{Corresponding Author}
\cortext[cor2]{Equally Contributed}

\begin{abstract}
In drug-resistant epilepsy, Stereo-Electroencephalography (SEEG) monitoring is one of the most effective techniques to identify the Epileptogenic Zone (EZ), the fundamental prerequisite for epilepsy surgery.
\begin{tikzpicture}[remember picture, overlay]
      \node[draw,minimum width=3.in] at ([yshift=-1cm]current page.north)  {This manuscript has been accepted for publication in \textit{Biomedical Signal Processing and Control}.};
\end{tikzpicture}
Despite recent technological advances, SEEG recordings remain difficult to interpret, and SEEG-guided surgery still achieves success rates below $70$\%.
In this work, we develop a novel computational framework for SEEG analysis, with the ultimate aim of improving the accuracy in EZ definition. 
Specifically, we investigate the hypothesis that epileptogenic regions exhibit a tendency to behave independently and thus \emph{desynchronize} from neighboring brain structures before seizure onset.
To this end, we design the Desynchronization Index (DI), an algorithm that identifies the Epileptogenic Zone (EZ) as the subset of channels that disconnect from the SEEG network during the ictal transition. 
We evaluate the DI algorithm against Epileptogenicity Index (EI), one of the most common tools for EZ definition, on a clinical dataset of $20$ patients, considering the channels that were thermocoagulated at the end of SEEG monitoring as the detection target.
Our results show that DI overcomes EI in terms of area under the ROC curve (AUC$=0.86$ vs. AUC$=0.83$), while combining the two algorithms into a single framework leads to the best performance (AUC$=0.88$).
Overall, the DI algorithm underscores anomalous connectivity patterns that are difficult to detect through visual inspection, improving the accuracy in the EZ definition and providing new insights into the dynamics of seizure generation.
\end{abstract}




\end{frontmatter}

\section{Introduction}

\label{sec:introduction}

The localization of the \gls{ez} is the fundamental prerequisite for performing epilepsy surgery in people with drug-resistant epilepsy~\cite{luders2006epileptogenic}.
Maximizing accuracy in the definition of the \gls{ez} makes it possible to reduce the extent of the surgery, mitigating the risk of complications or chronic deficits.
In complex scenarios, the \gls{ez} can only be identified through invasive procedures, among which \gls{seeg} represents the most effective solution~\cite{isnard2018french}.
This procedure involves the surgical implantation of electrodes into the patient's brain, enabling the recording of electrical activity from both superficial and deep cortical regions.
As a result, \gls{seeg}-guided surgical resections lead more than $60\%$ of patients to be seizure-free after the intervention~\cite{cossu2005stereoelectroencephalography}. 
However, \gls{seeg} provides very focused information: if no intracranial electrodes explore the \gls{ez}, the procedure may result in unsuccessful epilepsy surgery~\cite{cardinale2013stereoelectroencephalography}.  

Nowadays, \gls{seeg} interpretation is based mainly on visual analysis, focusing on the early phases of epileptic discharge.
The goal is to define an \gls{ez} that can be focal or distributed among multiple cortical structures, a scenario that falls under the concept of epileptogenic networks~\cite{bartolomei2017defining}.
This process is particularly complex, since the \gls{seeg} signals include hundreds of components, named \emph{channels}, which must be manually reviewed by highly specialized neurophysiologists.
In recent years, \gls{seeg} interpretation has being supported by computational tools that aim to characterize the \gls{ez} before and during seizure generation~\cite{gnatkovsky2014biomarkers}.
The most common methods extract features from the \gls{seeg} power spectrum, searching for the cortical sites generating \emph{fast oscillations}~\cite{7458827}, i.e., electrical activities with frequencies in the gamma range ($[30, 70]$ Hz), as done in the cases of \gls{ei}~\cite{bartolomei2008epileptogenicity} and \glspl{em}~\cite{david2011imaging}.

Complementary methodologies characterize \gls{seeg} signals according to the connectivity between the different channels.
In this context, the first studies analyzed the responses generated through intracranial stimulations, known as \glspl{ccep}~\cite{8532306}, which, however, do not have a physiological origin.
Other solutions analyze the phase distribution of the signals associated with each cortical site, as performed by the \gls{plv} and the \gls{pli}, evaluating the synchrony between the \gls{seeg} channels.
To estimate the direction of neural connections, we need more advanced methodologies, such as the \gls{pte} which, notably, was applied to \gls{seeg} data only in a limited number of cases~\cite{wang2017identification}.

Although several connectivity approaches have been developed for the study of epileptogenic networks, the literature often presents inconsistent results.
Pioneering investigations show that the \emph{ictal phase} (i.e., the period involving epileptic discharges) is preceded by a reduction in cortical connectivity, followed by the opposite phenomenon during seizure propagation~\cite{proix2017individual}.
In recent years, it has been hypothesized that the \gls{ez} presents an abnormal inward information flow during the inter-ictal phase~\cite{gunnarsdottir2022source}, and that the epileptogenic level of the cortical sites can be assessed by the strength of connections exerted after the seizure onset~\cite{nithin2021graph}.
Notably, most recent studies follow the latter approach and characterize \gls{seeg} signals during the diffusion of epileptic discharges.

The aim of this work is to investigate whether the ability of a brain signal to evolve independently of physiological activity can serve as a key marker of epileptogenicity.
From a \gls{seeg} perspective, this implies that epileptogenic channels become functionally disconnected from the rest of the \gls{seeg} network during the ictal transition.
This idea is supported by prior findings~\cite{bartolomei2001neural}, but, to our knowledge, has not yet been applied to quantify the epileptogenicity of individual \gls{seeg} channels. 
To validate our central hypothesis, we design a novel framework to support neurophysiologists in \gls{seeg} interpretation and to improve the definition of the \gls{ez}.
The core of the framework is a new connectivity-based algorithm, termed \gls{di}, which identifies the \gls{ez} as the subset of \gls{seeg} channels responsible for connectivity disruptions at the seizure onset.

Overall, our main contributions consist of the following points:
\begin{itemize}
    \item we design a new connectivity-based metric, termed \emph{desynchronization}, describing the tendency for a \gls{seeg} channel to self-isolate from the rest of the \gls{seeg}-explored brain network;
    \item we design a new algorithm, named \gls{di}, which estimates the degree of epileptogenicity of cortical sites according to the increase in desynchronization during the ictal transition;
    \item we implement the \gls{di} algorithm to analyze seizure data in $20$ patients with drug-resistant epilepsy who underwent \gls{seeg} monitoring at the IRCCS Institute of Neurological Sciences of Bologna (Italy);
    \item we analyze the performance of our method both against and in combination of the \gls{ei} algorithm, assessing its ability to identify the \gls{seeg} channels labeled as part of the \gls{ez} during clinical evaluation.
\end{itemize}


The manuscript is organized as follows:
Sec.~\ref{sec:related} reports the most relevant computational tools used for \gls{seeg} interpretation;
Sec.~\ref{sec:model} presents the mathematical model used to represent \gls{seeg} signals and the \gls{ei} algorithm;
Sec.~\ref{sec:method} describes the proposed \gls{di} algorithm and our evaluation methodology;
Sec.~\ref{sec:result} presents the results of the work;
Sec.~\ref{sec:discuss} discusses our findings from a clinical point of view;
Sec.~\ref{sec:concl} gives the conclusions and depicts some avenues for future work. 

\section{Related Work}
\label{sec:related}

In clinical practice, the interpretation of \gls{seeg} recordings constitutes a highly time-consuming process due to the huge number of data that must be visually reviewed by neurophysiologists~\cite{bartolomei2018interpretation}. 
The most common marker of the seizure onset is given by low voltage fast oscillations, which are electrical discharges spanning beta frequencies ($[13, 30]$ Hz), usually observed in mesial temporal seizures, and gamma frequencies ($[30, 100]$ Hz), observed in neocortical seizures.
To quantify the magnitude of fast oscillations and, more generally, the dynamics that lead to seizures, it is common to analyze the spectrum of \gls{seeg} signals.
This approach is followed by the \gls{ei} algorithm~\cite{bartolomei2008epileptogenicity}, which is the most routinely method to assist clinicians in \gls{seeg} interpretation~\cite{makhalova2023role}.

The \gls{ei} algorithm, occasionally combined with other biomarkers~\cite{balatskaya2020connectivity}, has been recognized to provide the best accuracy in terms of \gls{ez} localization and surgical prediction.
An extension of such an approach is the \emph{epileptogenicity rank}, a technique that, before computing the \gls{ei} values, assigns weights to the \gls{seeg} sites depending on their distance from the hypothesized target of the \gls{ez}~\cite{parasuram2021quantification}.
However, this solution biases the epileptogenic levels assigned to each channel by a prior assumption.
In particular, the epileptogenicity rank reduces the number of false positives when the clinical hypothesis is correct, but dramatically decreases the sensitivity in the other case. 

While \gls{ei} focus on the energy distribution in the frequency domain, other approaches analyze the variation in signal synchronization.
Several tools have been proposed for this purpose, including the \gls{plv}~\cite{lachaux1999measuring} and the \gls{pli}~\cite{stam2007phase}, with different trade-offs in terms of sensitivity and precision~\cite{gupta2020current}.  
Notably, the \gls{pli} was proposed as an alternative to the \gls{plv} in order to mitigate the noisy information associated with \emph{volume conduction}~\cite{peraza2012volume}.
A more advanced method is the \gls{pte}~\cite{lobier2014phase}, which allows us to estimate both the delay and the direction of neural connections, representing a promising yet underutilized technique for \gls{seeg} analysis~\cite{wang2017identification}.

In one of the first works in this field, Bartolomei et al. exploit the nonlinear regression coefficient (h2)~\cite{wendling2000relevance} and show that the \gls{ez} is associated with reduced connectivity at the beginning of the ictal phase~\cite{bartolomei2001neural}.
On the other hand, recent literature primarily characterizes the \gls{ez} after the seizure onset, with the epileptogenic levels of \gls{seeg} channels quantified in terms of functional connectivity during the ictal phase~\cite{huang2023intracranial,daoud2022stereo}.
Two examples are provided in~\cite{bernabei2021electrocorticography}, where functional connectivity is integrated with the Euclidean distance, and~\cite{li2021hubs}, where graph theory is used to identify network hubs during seizure propagation. 

More advanced connectivity-based approaches associate the phase of slow oscillations with the amplitude of fast oscillations, as occurred for the Phase Slope Index (PSI) and the Phase Amplitude Coupling (PAC) methods~\cite{bastos2016tutorial}.
In this regard, the authors of~\cite{jiang2022interictal} exploit PSI to analyze \gls{seeg} signals and confirm that epileptogenic channels show increased outward connectivity during the ictal phase.
Similar observations are presented in~\cite{liu2021epileptogenic}, whose authors computed the PAC between the high (gamma) and low (delta and theta) frequency bands. 
However, a recent work~\cite{an2020localization}, using Partial Direct Coherence (PDC)~\cite{baccala2001partial} as a connectivity measure, contradicts these findings and reports that the \gls{ez} presents greater inward connectivity during seizure propagation.


Despite the large variety of works that use connectivity-based metrics for the interpretation of \gls{seeg} signals, the results in the literature do not lead to an agreement on how to characterize the \gls{ez}.
Initial studies suggest that the ictal transition is characterized by a reduction in connectivity, which is consistent with the desynchronization hypothesis proposed in this work.
In contrast, more recent approaches identify the \gls{ez} as the subset of cortical sites exhibiting stronger coupling during seizure propagation. 
These discrepancies may depend on the specific filtering pipeline used to analyze \gls{seeg} data, as epileptogenic signals may lead to higher functional connectivity in high-frequency bands while reducing it within physiological frequency ranges.

\section{Mathematical Preliminaries}
\label{sec:model}

In this section, we first describe how single \gls{seeg} channels are modeled within our framework and define the \emph{energy ratio} as a baseline epileptogenicity biomarker. 
Then, we formulate the connectivity model used to infer the relation between \gls{seeg} sites starting from their phase distribution.
Finally, we introduce the \gls{ei} algorithm, which represents the state-of-the-art for defining the \gls{ez} and will serve as a benchmark for the evaluation of our computational framework.

\subsection{Channel Model}
\label{sec:channel}

We model an \gls{seeg} recording as a multidimensional signal with multiple channels, one for each cortical site analyzed. 
In the rest of the paper, we denote by $\mathcal{N}$ the set of channels and by $|\cdot|$ the cardinality operator, so that the number of cortical sites explored by the \gls{seeg} implant is $N=|\mathcal{N}|$.
In our model, each channel $x \in \mathcal{N}$ is segmented in multiple overlapping windows $x(t)$, with $t=0, \Delta t, 2 \Delta t,\text{...}$, where $\Delta t$ represents the time-shift between consecutive windows.
In particular, given the duration $T_{\text{window}}$ of a single window and the sampling frequency $f_s$, the number of samples per window is $n=T_{\text{window}} \cdot f_s$.

Baseline approaches to quantifying the epileptogenicity of \gls{seeg} channels rely on analyzing the distribution of signal energy in the frequency domain.
To obtain a frequency-domain representation of $x(t)$, we apply the \gls{ft}, obtaining a complex value $X(t,f)$ for each frequency in $[0, f_s / 2]$. 
Following the approach proposed in~~\cite{bartolomei2008epileptogenicity}, we then compute the \emph{energy ratio} between the high- and low-frequency bands of $x(t)$ as
\begin{equation}
\label{eq:energy_ratio}
    E_x(t) = \frac{ \int_{B_{h}} \lVert  X(t,f) \rVert^2 df}{ \int_{B_{\ell}}  \lVert  X(t,f) \rVert ^2 df},
\end{equation}
where $B_{h}$ and $B_{\ell}$ are the high- and low-frequency ranges, while $\lVert  \cdot \rVert$ is the norm function.
The straightforward idea behind the energy ratio is that epileptic discharge is characterized by an increase in fast oscillations and a concurrent decrease in slow oscillations. 
By normalizing the high-frequency energy by the low-frequency energy, we can fairly compare the signals recorded at different cortical sites, which may be characterized by different total energy. 

\subsection{Connectivity Model}
\label{sec:connection}

To estimate the relations between different \gls{seeg} channels, we consider the \gls{pte}, which models the coupling of any pair of channels $x, y \in \mathcal{N}$ according to their phase distributions.
Given a specific window $x(t)$, we first calculate its \emph{analytic representation} $x_a(t) = x(t) + j HT\left(x(t)\right)$, where $HT(\cdot)$ is the Hilbert transform.
We observe that $x_a(t)$ is associated with $n=T_{\text{window}} \cdot f_s$ phase values, which constitute the \emph{instantaneous phase} $\theta_x(t) \in [-\pi, +\pi]$ of channel $x$ at time $t$ and provide information about the signal’s cycle position independently of its amplitude.
Notably, in this manuscript, we do not apply any filtering before the Hilbert transform, so $\theta_x(t)$ represents the phase of the broadband signal associated with channel $x$.
Extending the approach to specific frequency ranges is also possible and will be investigated in future work.

To model the phase distribution, we use an histogram with uniform bin widths.
Hence, we follow the Sturges rule~\cite{freedman1981histogram} and compute the width of each bin as $\vartheta = 2\pi / (\log_2(n) + 1)$.
We observe that the width of each bin increases with the domain of the target variable, which, in the case of a phase, is equal to $2\pi$, and decreases with the number of samples, which in our model is fixed and equal to $n$.
This quantization step reduces the sensitivity to zero-lag coupling, attenuating the biases associated with volume conduction
\footnote{Another approach designed to mitigate volume conduction is the \gls{pli} algorithm, which evaluates neural synchronization by considering only the sign of phase differences.
However, \gls{pli} does not allow us to measure the direction of the information flows between signals~\cite{stam2007phase} and underestimates neural connections with phase difference that fluctuates around zero. 
The opposite approach is taken by \gls{plv}, which models the phase in a continuous domain, resulting more sensitive to both genuine and spurious zero-lag coupling connections.}.

Let us denote by $\theta_x(t)$ and $\theta_y(t)$ the phase distribution of $x(t)$ and $y(t)$, respectively. 
According to the \gls{pte}, the strength of the connections that channel $x$ exerts on channel $y$ at time $t$, considering a lag $\tau$, is given by
\begin{equation}
\begin{gathered}
    T_{x \rightarrow y} (t, \tau) = H\left(\theta_y(t), \theta_y(t + \tau)\right) +\\
    H\left(\theta_x(t), \theta_y(t)\right) - H\left(\theta_x(t), \theta_y(t), \theta_y( t + \tau) \right) - H\left(\theta_y(t)\right),
\end{gathered}
\label{eq:pte_lag}
\end{equation}
where $H(\cdot)$ denotes the Shannon entropy function.
Intuitively, $T_{x \rightarrow y} (t, \tau)$ represents the amount of information that $\theta_y(t + \tau)$ shares with $\theta_x(t)$ and that cannot be predicted from the past of values of $\theta_y$.

We observe that the \gls{pte} is a directed connectivity measure, which means that, in general, $T_{x \rightarrow y} (t, \tau) \neq T_{y \rightarrow x} (t, \tau)$.
In order to remove the dependency from $\tau$, with a slight abuse of notation, we redefine the \gls{pte} between $x(t)$ and $y(t)$ as the maximum value of $T_{x \rightarrow y} (t, \tau)$ among multiple lags in the set $ [0,  \tau _{max}]$:
\begin{equation}
    T_{x \rightarrow y} (t) = \max_{  \tau \in [0,  \tau _{max}]} T_{x \rightarrow y} (t, \tau).
\end{equation}
By doing so, we obtain that the magnitude of the effective connection exerted on channel $y$ by channel $x$ at time $t$ is given by $T_{x \rightarrow y} (t)$, while the propagation delay associated with such a connection is:
\begin{equation}
    \tau_{x \rightarrow y} (t) = \argmax_{  \tau \in [0,  \tau _{max}]} T_{x \rightarrow y} (t, \tau).
\end{equation}
We observe that if $\tau_{x \rightarrow y} (t) = 0$, there is a zero propagation delay for the information flow going from $x(t)$ to $y(t)$.

\subsection{Epileptogenicity Index}
\label{sec:epil}

The \gls{ei} algorithm considers the energy ratio as a biomarker for identifying abnormal channels during the ictal transition and, consequently, defining the \gls{ez}.
However, \gls{ei} is not designed to simply assign higher epileptogenicity values to channels with higher energy ratios, as high-frequency oscillations can occur even in the absence of epileptic discharges.
To address this issue, the \gls{ei} algorithm implements a \gls{cusum} control chart to discern physiological from epileptogenic signals.
Given a sequence of observations $\omega(t)$, interspersed by a period $\Delta t$, the \gls{cusum} chart is defined by the function
\begin{equation}
\label{eq:cusum}
    \Gamma(\omega, t) =
    \begin{cases}
        \max \left\{ 0,  \Gamma(\omega, t-\Delta t) + \frac{\omega(t) - \mu_{\omega}}{\sigma_{\omega}} - \gamma \right\},  & t > 0; \\
        \omega(t),  & t = 0;
    \end{cases}
\end{equation}
where $\gamma \in \mathbb{R}$ is a tuning parameter that makes the statistic less or more sensitive to the new observations, while $\mu_{\omega}$ and $\sigma_{\omega}$
are the estimates of the mean and standard deviation of $\omega$.
In the case of the \gls{ei} algorithm, the observations $\omega(t)$ are given by the energy ratios $E_x(t)$ associated the \gls{seeg} channels $x \in \mathcal{N}$.

The original version of the \gls{ei} algorithm does not consider the standard deviation in the normalization process, and dynamically re-estimates the mean energy ratio every time the energy ratio varies significantly.
In this manuscript, we estimate the baseline statistic of each channel $x \in \mathcal{N}$  by looking at the period immediately preceding the ictal discharge.
In particular, we compute $\mu_{E_x}$ and $\sigma_{E_x}$ over the interval $\{t_{\text{base}}, ..., t_{\text{start}}\}$, where $t_{\text{start}}$ is the instant that corresponds to the seizure onset.
Then, we compute the cumulative sum $\Gamma(E_x, t)$ of the energy ratio $E_x(t)$ for each time $t \in \{t_{\text{start}}, ..., t_{\text{end}}\}$, where $t_{\text{end}}$ is the instant at which epileptic discharge have propagated within the whole brain.

Given a channel $x \in \mathcal{N}$ and $\Gamma(E_x, t)$, we define the \emph{activation time} as the instant of time at which $\Gamma(E_x, t)$ reaches the highest value:
\begin{equation}
\label{eq:activation}
    t_{E_x} = \argmax_{t \in \{t_{\text{start}}, ..., t_{\text{end}}\}} \Gamma (E_x, t)
\end{equation}
Hence, we define the \emph{cumulative energy} of the channel as the sum of the energy ratio $E_x$ over the period following the channel's activation time:
\begin{equation}
\label{eq:tonicity}
    c_{E_x} = \sum_{t=t_{E_x}}^{t_{E_x} +\, m \cdot \Delta t} E_x(t),
\end{equation}
where $m \in \mathbb{Z}^+$ is the number of windows over which $c_{E_x}$ is calculated.
Notably, the latter coincides with the integral of the energy ratio over a given time period and represents the magnitude of the abnormality of channel $x$, quantified in terms of the increase of high-frequency oscillations.

Finally, \gls{ei} identifies the \gls{ez} as the group $\mathcal{N}_{E}$ of channels with the strongest variations in the energy ratio, that is
\begin{equation}
   \mathcal{N}_{E} = \left\{ x \in \mathcal{N}: \Gamma(E_x, t_{E_x}) > \eta \cdot \max_{y, t} \Gamma(E_y, t) \right\},
\end{equation}
where $\eta \in [0, 1]$ is the \emph{detection threshold}. 
The epileptogenicity level of each channel $x \in \mathcal{N}$ is $EI_x = 0$, $\forall$ $x \in \mathcal{N}\, \backslash \, \mathcal{N}_{E}$, and 
\begin{equation}
    EI_x = \frac{c_{E_x}}{t_{E_x} - t_{\text{start}}}, \, \forall x \in \mathcal{N}_E.
\end{equation}
We observe that $EI_x$ is proportional to $c_{E_x}$ and decreases as a function of the difference between $t_{E_x}$ and $t_{\text{start}}$, i.e., the time delay with which channel $x$ begins to exhibit high-frequency oscillations relative to seizure onset.

We note that our formulation stands out from the original \gls{ei} algorithm by reducing the number of parameters to be manually set.
The times $t_{\text{start}}$ and $t_{\text{end}}$ denote the period during which epileptic discharges are generated, while $t_{\text{base}}$ must ensure that $\mu_{E_x}$ and $\sigma_{E_x}$ capture the baseline brain activity of the patient. 
The only parameters to be manually tuned are the number $m$ of windows over which $c_{E_x}$ is computed, the weight $\gamma$ of the \gls{cusum} control chart, and the threshold $\eta$ which, notably, trades off false alarms and miss-detections.

\section{Methodology}
\label{sec:method}

\glsreset{di}

In this section, we present our novel detection algorithm, the \gls{di}, which is designed to quantify the epileptogenic level of \gls{seeg} channels based on their tendency to self-isolate from the \gls{seeg} network during the ictal transition.
We then describe the pipeline used to evaluate our approach, including the \gls{seeg} recording process, the clinical dataset, and the metrics used to assess the performance of the overall detection framework. 

\subsection{Desynchronization Index}
\label{sec:desync}

The \gls{di} algorithm is built on a novel epileptogenicity biomarker, termed \emph{desynchronization}, which characterizes epileptogenic channels according to their reduced connectivity with respect to the rest of the \gls{seeg} network. 
To design this biomarker, we consider a similar approach to~\cite{sparks2019monitoring}, where the goal is to detect anomalous nodes in time-varying networks. 

Following the model in Sec.~\ref{sec:connection}, we write $\mathcal{T}(t)$ to indicate the distribution of connections $T_{x \rightarrow y}(t)$, $\forall \, x, y \in \mathcal{N}$, in the \gls{seeg} network at time $t$, and $\mathcal{T}_{n}(t)$ to denote the $n$-th percentile of the distribution.
Hence, to separate genuine from spurious interactions, we define $\mathcal{N}_x^{\text{low}} (t)$ and $\mathcal{N}_x^{\text{high}} (t)$ as the sets of channels that exhibit significantly reduced and increased information flow originating from channel $x$:
\begin{align}
    \mathcal{N}_x^{\text{low}} (t) = & \big\{ y \in \mathcal{N}: T_{x \rightarrow y} (t)  \leq  \mathcal{T}_{n_{\text{low}}}(t) \big\},\\
    \mathcal{N}_x^{\text{high}} (t) = & \big\{ y \in \mathcal{N}: T_{x \rightarrow y} (t)  \geq  \mathcal{T}_{n_{\text{high}}}(t) \big\}.
\end{align}

In other words, a connection is deemed significant if its strength lies below the $n_{\text{low}}$-th or exceeds the $n_{\text{high}}$-th percentile of the time-varying network distribution.
This allows us to implicitly customize the framework to each patient without the need to define additional hyperparameters, which is a strong limitation in~\cite{sparks2019monitoring}.
We observe that other rules for discerning significant connections could lead to different trade-offs in terms of accuracy, and further investigation of how to tune $n_{\text{low}}$ and $n_{\text{high}}$ will be part of our future work. 

We now define the actual $\psi_x^{\text{low}}(t)$ and awaited $\hat{\psi}_x^{\text{low}}(t)$ densities of low-strength connections originating from channel $x$ as
\begin{equation}
    \psi_x^{\text{low}}(t) = \sum_{y \in \mathcal{N}_x^{\text{low}} (t)} T_{x \rightarrow y} (t),
\end{equation}
\begin{equation}
    \hat{\psi}_x^{\text{low}}(t) = |\mathcal{N}_x^{\text{low}} (t)| \cdot \mathcal{T}_{n_{\text{low}}}(t).
\end{equation}
Similarly, we define the actual $\psi_x^{\text{high}}(t)$ and awaited $\hat{\psi}_x^{\text{high}}(t)$ densities of high-strength connections originating from channel $x$ as
\begin{equation}
    \psi_x^{\text{high}}(t) = \sum_{y \in \mathcal{N}_x^{\text{high}} (t)} T_{x \rightarrow y} (t),
\end{equation}
\begin{equation}
    \hat{\psi}_x^{\text{high}}(t) = |\mathcal{N}_x^{\text{high}} (t)| \cdot \mathcal{T}_{n_{\text{high}}}(t).
\end{equation}
Hence, $\psi_x^{\text{low}}(t)$ is proportional to the cumulative magnitude of low-strength outward connections of channel $x$, while $\psi_x^{\text{high}}(t)$ is proportional to the cumulative magnitude of high-strength outward connections of the same channel. By definition, we always have $\psi_x^{\text{low}}(t) < \hat{\psi}_x^{\text{low}}(t)$ and $\psi_x^{\text{high}}(t) > \hat{\psi}_x^{\text{high}}(t)$.

We define the desynchronization $D_x(t)$ of channel $x \in \mathcal{N}$ at time $t$ as
\begin{equation}
\label{eq:desync_level}
    D_x(t) = \frac{\sqrt{\hat{\psi}_x^{\text{low}}(t)} - \sqrt{\psi_x^{\text{low}}(t)}}{\epsilon + \sqrt{\psi_x^{\text{high}}(t)} - \sqrt{\hat{\psi}_x^{\text{high}}(t)}},
\end{equation}
where the square root operator is used to stabilize the variance of the process~\cite{sparks2019monitoring}, and $\epsilon>0$ is a small constant to avoid numerical stability in edge cases.
The denominator of the above equation decreases as $x$ presents a lower number of high-strength outward connections, while the numerator increases as $x$ presents a higher number of low-strength outward connections. 
Therefore, $D_x(t)$ represents the tendency of $x$ to self-isolate from the rest of the network. 

We observe that both the denominator and the numerator of Eq.~\eqref{eq:desync_level} always take positive values and, consequently, $D_x(t) \in \mathbb{R}^+$, $\forall$ $x$ $\in \mathcal{N}$.
In the edge case, a node $x$ may present only null-value connections, which leads $\hat{\psi}_x^{\text{high}}(t)=\psi_x^{\text{high}}(t)=\psi_x^{\text{low}}(t)=0$, $\hat{\psi}_x^{\text{low}}(t)=N \cdot \mathcal{T}_{n_{\text{high}}}(t)>0$, and, consequently, $D_x(t) \rightarrow + \infty$.
This numerical instability is avoided by the constant $\epsilon > 0$ in the denominator, ensuring that $D_x(t)$ is upper bounded by $(N \cdot \mathcal{T}_{n_{\text{high}}}(t))^{1/2} \cdot \epsilon^{-1} \leq (N \cdot T_{\text{max}})^{1/2} \cdot \epsilon^{-1} $, where $T_{\text{max}}$ is the maximum \gls{pte} value.

To identify epileptogenic channels, \gls{di} follows the same approach as \gls{ei}, monitoring which channels exhibit the most prominent desynchronization changes during the ictal transition.
Using the pipeline presented in Sec.~\ref{sec:epil}, we apply the \gls{cusum} chart and obtain $\Gamma(D_x, t)$ for each channel $x \in \mathcal{N}$ and for each time $t$ between $t_{\text{start}}$ and $t_{\text{end}}$.
Hence, using equations \eqref{eq:activation} and \eqref{eq:tonicity}, we compute the activation times $t_{D_x}$ and the cumulative desynchronization $c_{D_x}$, $\forall$ $x \in \mathcal{N}$. 
Finally, we define $\mathcal{N}_D$ as the set of channels that show the strongest changes in terms of desynchronization, and compute the \gls{di} value of each channel $x \in \mathcal{N}_D$ as the ratio between $c_{D_x}$ and the delay of channel's activation:
\begin{equation}
   \mathcal{N}_{D} = \left\{ x \in \mathcal{N}: \Gamma(D_x, t_{D_x}) > \eta \cdot \max_{y, t} \Gamma(D_y, t) \right\},
\end{equation}
\begin{equation}
\label{eq:desync_index}
    DI_x =
    \begin{cases}
     \frac{c_{D_x}}{t_{D_x} - t_{\text{start}}} & x \in \mathcal{N}_D, \\
     0 & x \notin \mathcal{N}_D.
    \end{cases}
\end{equation}
The \gls{di} algorithm presents the same hyperparameters of \gls{ei}, making it necessary to tune only the threshold $\eta \in [0, 1]$, the number $m$ of windows to compute the cumulative desynchronization, and the weight $\gamma$ of the \gls{cusum} chart.

We observe that the threshold $\eta$ establishes the channels included in the \gls{ez}, i.e., the composition of $\mathcal{N}_{D}$, while $DI_x$ is used to assess the importance of the different channels within the \gls{ez} itself.
In the original work by Bartolomei and colleagues~\cite{bartolomei2008epileptogenicity}, an additional threshold was applied to the index values to distinguish propagator channels from truly epileptogenic ones.
In this work, we avoid introducing this additional detection level in order to enhance the explainability of the framework and to focus exclusively on determining which feature, energy ratio or desynchronization, is more suitable for defining the \gls{ez}.

\subsection{Clinical Dataset}
\label{sub:dataset}

This work considers an initial cohort of $24$ consecutive patients that have been monitored through \gls{seeg} at the IRCCS Institute of Neurological Sciences of Bologna from January 2022 to September 2025.
The study protocol was approved by the local ethics committee (protocol number 25058, committee code 97338), and written informed consent was obtained from each patient.

Before implantation, all patients underwent a comprehensive non invasive presurgical evaluation, including long-term video-EEG, 3T brain magnetic resonance imaging, neuro-psychological assessment, and FDG-PET.
The findings of these non-invasive examinations were then integrated to formulate the localization hypothesis for the \gls{ez}, which subsequently guided SEEG electrode planning.
The \gls{seeg} implantation followed the workflow developed at Niguarda Hospital~\cite{cardinale2015cerebral}, which involves the construction of a multimodal scene of the patient's brain.
The scene allowed for a comprehensive evaluation of all the anatomical information regarding the cortical area explored by each contact.

Each \gls{seeg} implant included multiple electrodes and each electrode had $5$–$18$ recording sites, corresponding to the channels of the recorded signal.
The number and location of the electrodes were tailored per patient, while each contact was $22$ mm in length, and separated by $1.5$ m from neighboring contacts.
The \gls{seeg} signals were recorded using the Nihon Kohden 2100 polygraph, considering 192 or 256 channels and a sampling frequency of $f_s=1000$ Hz.
High-definition synchronized videos were recorded throughout each \gls{seeg} monitoring (up to 20 days), enabling the correlation between electrical and clinical features. 

From the initial cohort, we excluded $4$ patients who did not achieve seizure control.
The final population comprised $11$ males and $9$ females with an average age of $34$ years (range $19$-$50$) at the time of recording, and $12$ years (range $1$-$37$) at the onset of the disease.
Clinical evaluation of the \gls{seeg} signals localized the \gls{ez} in the temporal lobe in $10$ cases ($7$ right, $3$ left), temporo-occipital region in $3$ cases ($2$ right, $1$ left), temporo-basal region in $2$ cases ($1$ right, $1$ left), insular lobe in $2$ cases ($1$ right, $1$ left), orbito-frontal region in $2$ cases ($1$ right, $1$ left), and right frontal lobe in $1$ case.

After one year of follow-up, seizure control was achieved in $13$ patients ($9$ Engel I, $4$ Engel II) who underwent \gls{rftc}, $6$ patients (Engel I) who underwent temporal lobectomy, and a single patient who underwent cortectomy (Engel II).
Histopathological analysis revealed findings compatible with Hippocampal Sclerosis (HS) in $4$ cases and Focal Cortical Dysplasia (FCD) in $2$ cases.
The demographic and clinical details of the overall population are reported in Tab.~\ref{tab:clinical}.

\begin{table*}[t!]
\scriptsize
\centering
\caption{Demographic and clinical data.}
\label{tab:clinical}
\begin{tabular}{cccccccccc}
  \toprule
    \shortstack{Patient \\ index} & \shortstack{Sex \\ \,} & \shortstack{Age \\(SEEG)} & \shortstack{Age \\(onset)} & \shortstack{Seizure \\ frequency} & \shortstack{Epileptogenic zone\\ localization} & \shortstack{Surgery\\ \,} & \shortstack{Outcome\\ (1 year) } & \shortstack{Histology\\ \,} \\
    \midrule 
    1 &   Male &          36 &            9 &            Weekly &                            Right temporo-occipital &                                            Cortectomy &    Engel II &               N/A \\
    2 & Female &          49 &           12 &           Monthly &                                      Left temporal &                                      Lobectomy &          Engel I &     HS 1 \\
    3 &   Male &          48 &            7 &            Annual &                                Right frontal &                                                      RF-TC &    Engel II &               N/A \\
    4 & Female &          25 &            2 &            Weekly &                                Right temporo-basal &                                 Lobectomy &          Engel I &                     Gliosis \\
    5 & Female &          26 &           12 &            Weekly &                                      Left temporal &                                    RF-TC &           Engel I &               N/A \\
    6 & Male &          39 &           33 &            Weekly &                                      Right temporal &                                   Lobectomy &           Engel I &               HS 1 \\
    7 & Female &          31 &           13 &             Daily &                               Right orbito-frontal &                         RF-TC &           Engel I &               N/A \\
    8 &   Male &          24 &            1 &             Daily &                                Left orbito-frontal &                                    RF-TC &           Engel I &            FCD \\
    9 &   Male &          32 &           16 &           Monthly &                                   Right temporal &                                      RF-TC &    Engel II &               N/A \\
    10 &   Male &          30 &           12 &            weekly &                                Left temporo-basal &                                      RF-TC &    Engel II &               N/A \\
    11 & Female &          34 &           22 &           Monthly &                              Left temporo-occipital &                                    RF-TC &           Engel I &               N/A \\
    12 & Female &          29 &            2 &             Daily &                                       Right insular &                             RF-TC &    Engel II &               N/A \\
    13 & Female &          43 &           37 &           Monthly &                               Right temporal &                                RF-TC &          Engel I &               N/A \\
    14 &   Male &          36 &           16 &           Monthly &                                      Left temporal &                  Lobectomy &     Engel I & FCD \\
    15 & Female &          50 &            3 &           Monthly &                               Right temporal &              Lobectomy &     Engel I &     HS 1 \\
    16 &   Male &          19 &            5 &           Monthly &                               Right temporal &                                   RF-TC &    Engel I &               N/A \\
    17 &   Male &          32 &           16 &             Daily &                                        Left insular &                                    RF-TC &          Engel I &               N/A \\
    18 & Female &          38 &           13 &           Monthly &                            Right temporo-occipital &                              RF-TC &          Engel I &               N/A \\
    19 &   Male &          24 &           18 &            Weekly &                                     Right temporal &                                  RF-TC &          Engel I &                       Tumor \\
    20 &   Male &          48 &           22 &            Weekly &                                     Right temporal &                  Lobectomy &          Engel I &     HS 1 \\
  \bottomrule
  \end{tabular}
\end{table*}

\subsection{Analysis Pipeline}
\label{sub:metric}
To build the signal dataset, we select a single seizure per patient and analyze an \gls{seeg} epoch of $T_{\text{epoch}}=200.0$ seconds for each seizure.
When multiple ictal events are available, we discard seizures recorded during the first two days of the monitoring period and those occurring within a seizure cluster, and select the first seizure among the remaining ones.
For each epoch, we set the start time so that the seizure onset occurs at $t_{\text{start}}=170.0$ seconds, allowing us to consistently study the ictal transition across the entire population.

Before running the \gls{ei} and \gls{di} algorithms, we process the \gls{seeg} signals through a comb filter centered at $f_{\text{comb}}=50$ Hz to suppress the powerline frequency and its harmonics.
In this phase, we remove \gls{seeg} channels that exhibit recording artifacts, including voltage saturation and other forms of signal distortion.
Notably, the labeling of \emph{bad channels} was performed by the clinicians during routine visual inspection of \gls{seeg} recordings.
No other cortical sites are excluded, ensuring that the proposed computational framework remains independent of both patient-specific and seizure-specific factors.

We recall that the outputs of the \gls{ei} and \gls{di} algorithms consist of the subset of \gls{seeg} channels classified as epileptogenic.
To establish a clinical ground truth, we selected channels that underwent \gls{rftc}, as in our population this procedure was applied to all channels clinically identified as part of the \gls{ez}.
Hence, for each epoch in the dataset, we implement the \gls{ei} and \gls{di} algorithms both as standalone tools and in a joint configuration. 
In the latter case (named “EI and DI”), a channel is classified as epileptogenic when at least one of the algorithms identifies it as positive. 
For a fixed detection threshold $\eta$, the “EI and DI” system is expected to improve sensitivity, at the cost of an increased number of false positives.

To assess the performance of the algorithms and their combination, we compute, for each epoch in the dataset, the following metrics:
\begin{itemize}
    \item \emph{Sensitivity} (or \emph{true positive rate}), which is the ratio between the number of channels correctly classified as epileptogenic by the algorithm and the total number of epileptogenic channels in the \gls{seeg} implant; 
    \item \emph{Precision} (or \emph{positive predictive value}), which is the ratio between the number of channels correctly classified as epileptogenic and the total number of channels classified as epileptogenic by the algorithm; 
    \item \emph{F1 score}, which is defined as the harmonic mean of precision and recall and provides a balanced measure of a model’s ability to correctly identify epileptogenic channels while accounting for false positives.
\end{itemize}
Furthermore, to study the trade-off between sensitivity and precision, we calculate the \gls{roc} and the \gls{auc}.
The \gls{auc} is a common performance indicator for diagnostic tools: in clinical settings, a value greater than $0.80$ is required to obtain a reliable solution~\cite{hosmer2013applied}. 
To ensure that the detection performance is not biased by the low number of positive elements, we also computed the \gls{pr-auc}.
This metric is particularly informative for imbalanced dataset, where the \gls{auc} can appear artificially high.

The computational framework used to process the \gls{seeg} data and obtain the results of this manuscript is publicly available at the link: \url{https://github.com/masonfed/desync\_index}.

\section{Results}
\label{sec:result}
In this section, we evaluate the performance of the \gls{ei} and \gls{di} algorithms to define the \gls{ez}, considering the analysis pipeline given in Sec.~\ref{sub:metric}.
In particular, we first conduct an exploratory analysis to assess how \gls{seeg} connectivity changes within the \gls{ez} during the ictal transition.
We then analyze in detail two representative seizure events, highlighting the differences between \gls{ei} and \gls{di} for detecting epileptogenic channels.
Finally, we examine the detection results aggregated across the entire patient population.

\subsection{Exploratory Analysis}
\label{sub:exploration}

Each \gls{seeg} epoch is segmented in overlapping time windows lasting $T_{\text{window}}=1.0$ s each, considering a time shift of $\Delta t = 0.25 $ s between consecutive windows.
Thus, given the epoch duration ($T_{\text{epoch}}=200.0$ s) a total of $797$ windows is analyzed for each seizure.
For computing the energy ratio, the high and low frequency ranges are set to $B_{h} = [30.0, 250.0]$ Hz and $B_{\ell} = [4.0, 14.0]$ Hz, respectively.
We thus extend the portion of the spectrum processed by the \gls{ei} algorithm, whose initial version considered frequencies lower than $97$ Hz~\cite{bartolomei2008epileptogenicity}.
By considering a wider frequency range, we can capture \emph{ripple} phenomena that are notably characterized by oscillations up to $250$ Hz or even more~\cite{urrestarazu2007}. 

For computing the desynchronization values, we consider $\tau_{max}=0.25$ s as the maximum propagation delay, and $n_{\text{low}}=5$ and $n_{\text{high}}=95$ to discern significant and spurious interactions in the \gls{seeg} network.
Hence, we compute $c_{E_x}$ and $c_{D_x}$ over $m=20$ windows, as done in~\cite{bartolomei2008epileptogenicity}, and we set the weight of the \gls{cusum} chart to $\gamma=0$, following the recommendations provided in~\cite{montgomery2019introduction}.
We set the start time of the baseline period to $t_{\text{base}} = t_{\text{start}} - 100.0$ s, and the end time of the detection procedure to $t_{\text{end}} = t_{\text{start}} + 20.0$ s. 
As explained in Sect.~\ref{sec:method}, $t_{\text{start}}$ coincides with the seizure onset and is equal to $170.0$ s.

\begin{figure}[h!]
     \centering
     \begin{subfigure}{.49\linewidth}
        \centering
        \includegraphics[width=\linewidth]{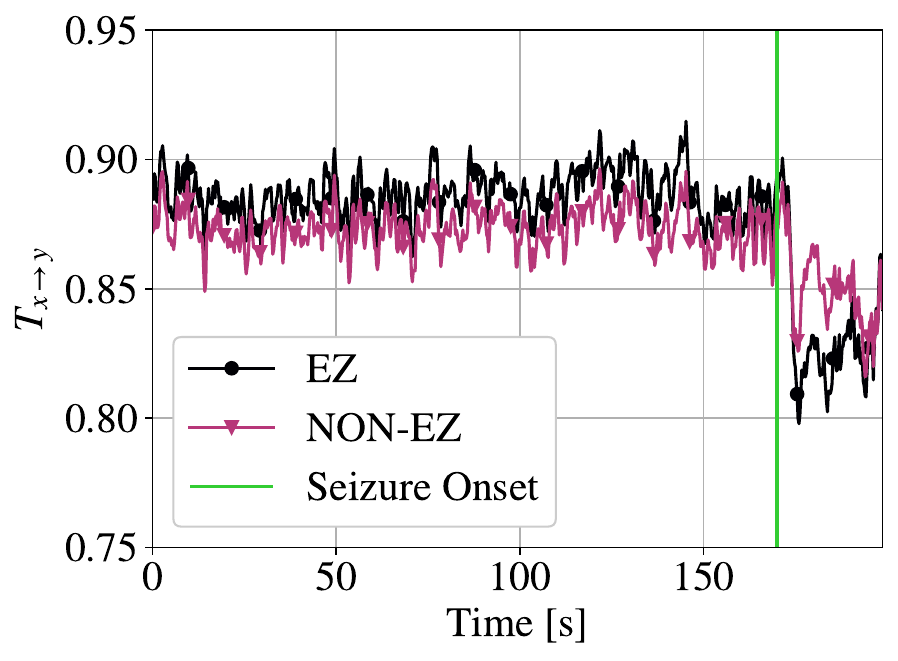}
        \caption{Connectivity in time.}
        \label{fig:outward_connection_series}
     \end{subfigure}
     \begin{subfigure}{.49\linewidth}
        \centering
        \includegraphics[width=\linewidth]{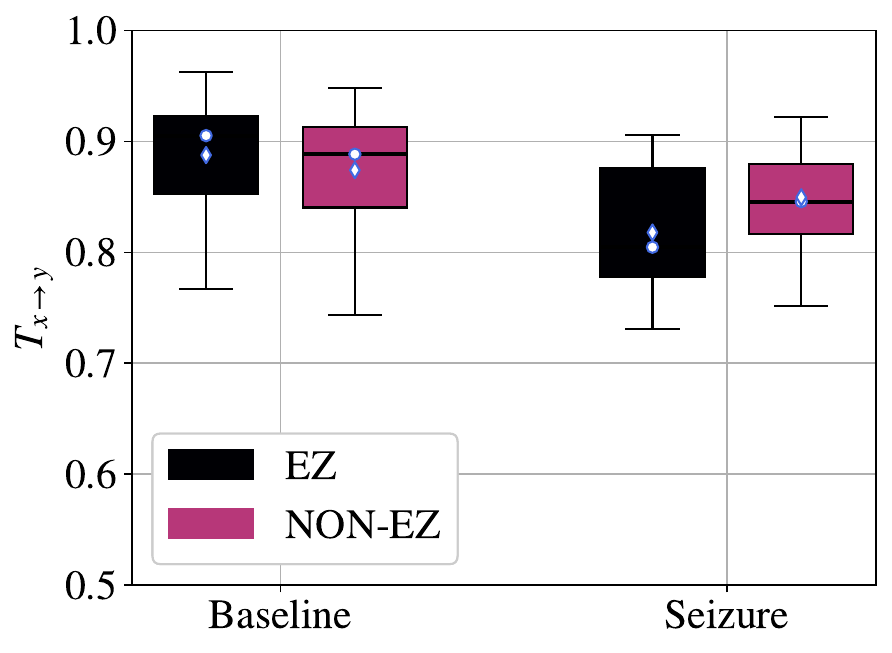}
        \caption{Connectivity distribution.}
        \label{fig:outward_connection_box}
     \end{subfigure}
     \caption{Outward connectivity during the ictal transition.}
     \label{fig:outward_connection}
\end{figure}

Before considering detection performance, we characterize the \gls{ez} and the rest of the \gls{seeg} network in terms of outward connectivity.
Fig.~\ref{fig:outward_connection_series} represents the mean outward connectivity for \gls{ez} and NON-\gls{ez} channels, over the $200.0$ seconds analyzed with the results averaged across the entire population. 
The \gls{ez} is associated with a greater outward connection than the rest of the network before the seizure onset ($t<t_{\text{start}})$.
Instead, after the start of epileptic discharges, both epileptogenic and non-epileptogenic channels present a drop in connectivity, which, in the case of the \gls{ez}, goes from $\approx0.9$ to $\approx0.8$.

In Fig.~\ref{fig:outward_connection_box}, we represent the distribution of the mean outward connectivity $T_{x \rightarrow y}$ of each channel $x$ within the \gls{ez} and NON-\gls{ez} before and after $t_{\text{start}}$. 
In particular, we consider the interval $\{t_{\text{base}}, ..., t_{\text{start}}\}$ as the baseline and the interval $\{t_{\text{start}} + 5.0 \text{ s}, ..., t_{\text{start}} + 15.0 \text{ s}\}$ as the seizure period. 
By performing a one-sided two-sample t-test ($\alpha=0.05$), we prove that this connectivity outage is strongly significant for the \gls{ez} ($\text{p-value}<0.001$) and marginally significant for the rest of the \gls{seeg} network ($\text{p-value}=0.063$). 
These results suggest that outward connectivity is particularly effective in identifying epileptogenic channels and support the use of the \gls{di} algorithm as a tool for precise \gls{ez} localization.

\subsection{Concordant Example}
\label{sub:concordant}

\begin{figure}[h!]
    \centering
    \begin{subfigure}{.49\linewidth}
        \centering
        \includegraphics[trim={0.425cm 0 0 0},clip,width=.99\linewidth]{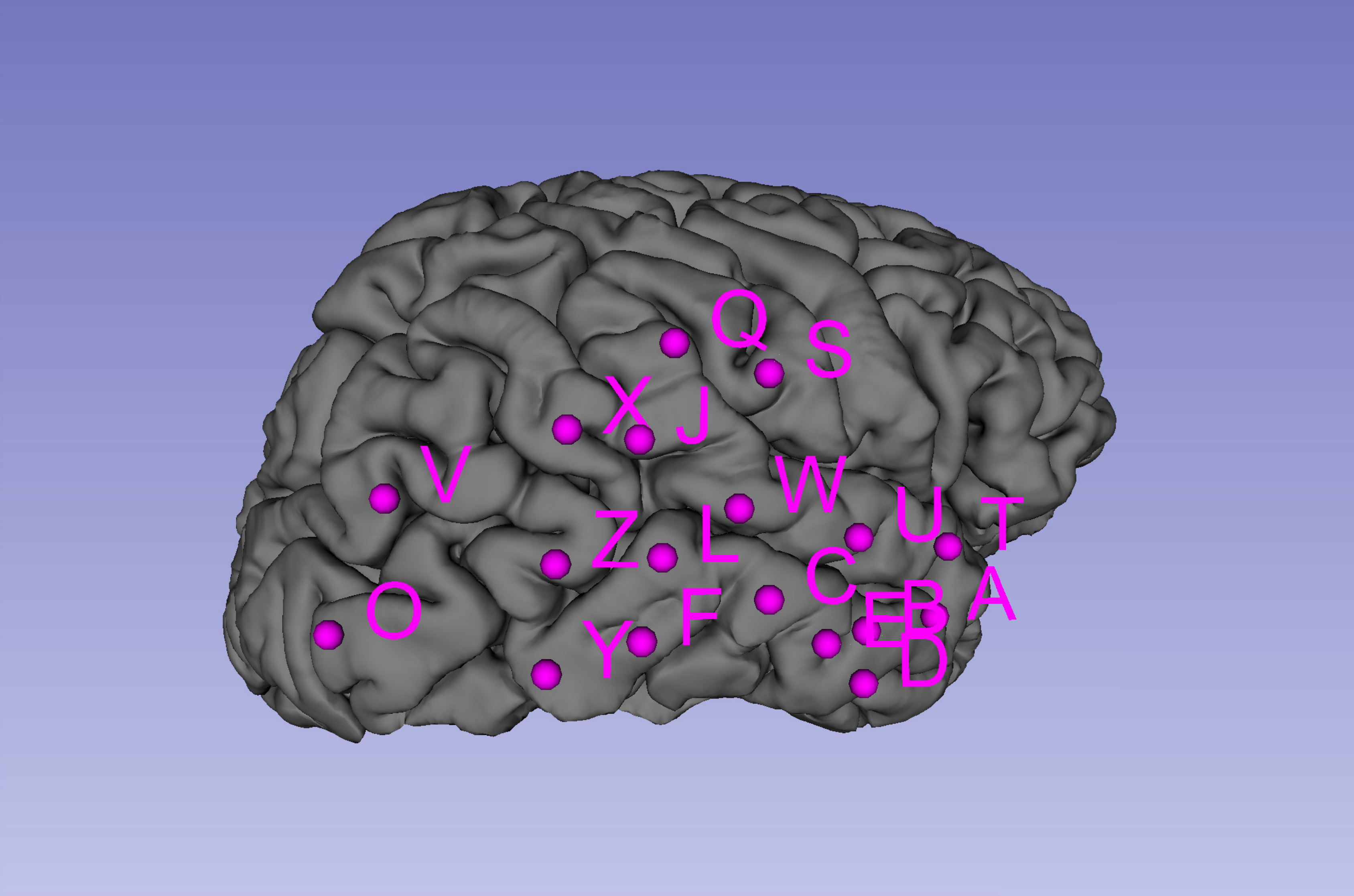}
        \caption{Patient $1$.}
        \label{fig:implant_1}
    \end{subfigure}
    \begin{subfigure}{.49\linewidth}
        \centering
        \includegraphics[width=.99\linewidth]{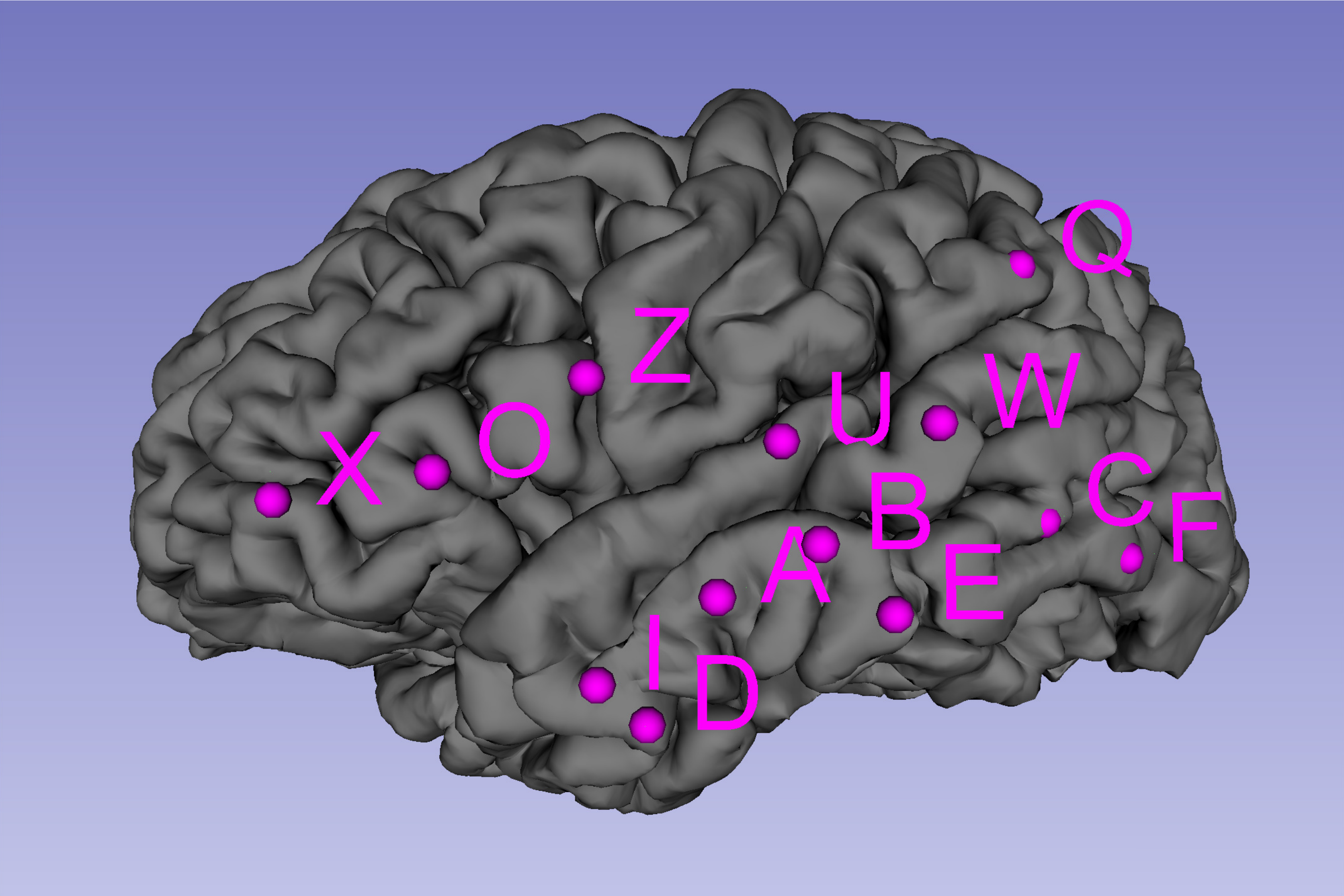}
        \caption{Patient $2$.}
        \label{fig:implant_2}
    \end{subfigure}
    \caption{SEEG implants.}
    \label{fig:implant}
\end{figure}

\begin{figure}[h!]
     \centering
     \begin{subfigure}{.49\linewidth}
         \centering
         \includegraphics[width=.99\linewidth]{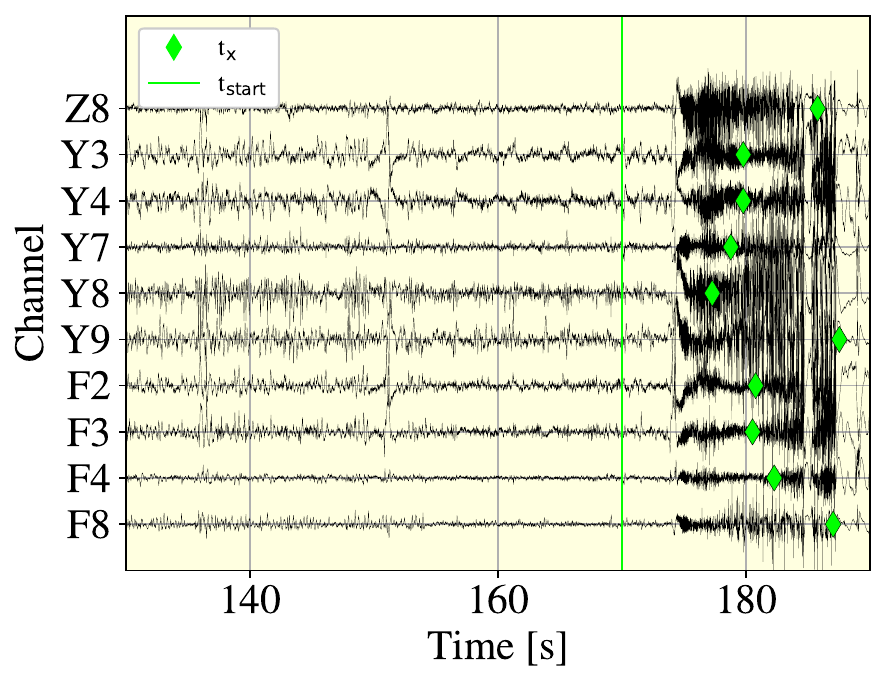}
         \caption{Signal.}
         \label{fig:energy_signal_concordant}
     \end{subfigure}
     \begin{subfigure}{.49\linewidth}
         \centering
         \includegraphics[width=.99\linewidth]{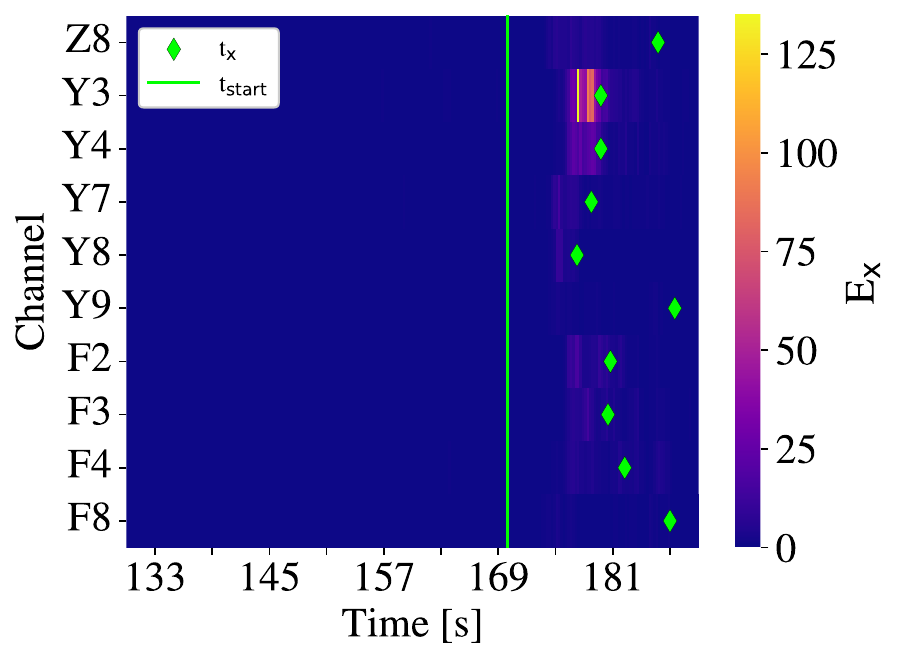}
         \caption{Energy ratio.}
         \label{fig:energy_concordant}
     \end{subfigure}
     \caption{The \gls{ei} algorithm (patient $1$, $\eta=16\%$).}
     \label{fig:ei_concordant}
\end{figure}

In the following, we focus on the operative characteristics of the \gls{ei} and \gls{di} algorithms, illustrating how each identifies \gls{ez} channels in two representative seizure events.
We first consider patient $1$, whose \gls{seeg} implant (represented in Fig.~\ref{fig:implant_1}) explores the right temporo-parietal and occipital areas, including $189$ different recording sites ($18$ electrodes).
Fig.~\ref{fig:energy_signal_concordant} reports the bipolar representation of the channels that are identified as epileptogenic by the \gls{ei} algorithm when setting $\eta=16\%$, while Fig.~\ref{fig:energy_concordant} represents the energy ratios associated with each of those channels.
In the two figures, $t_{\text{start}}$ denotes the seizure onset, while the markers $t_x$ denote the times at which the energy ratio $E_x$ diverges from its baseline. 

\begin{figure}[h!]
     \centering
     \begin{subfigure}{.49\linewidth}
         \centering
         \includegraphics[width=.99\linewidth]{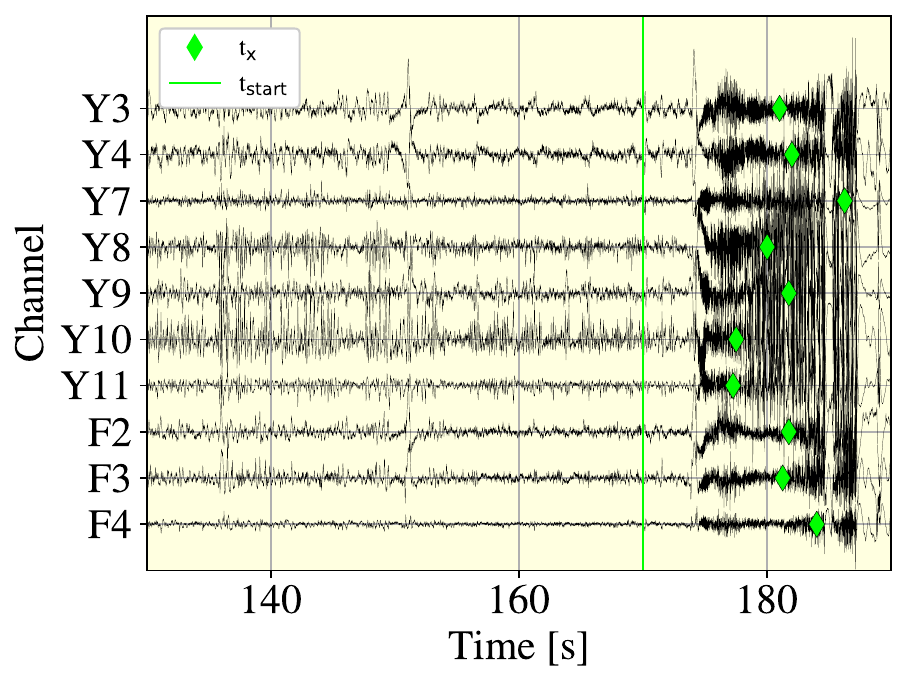}
         \caption{Signal.}
         \label{fig:connection_signal_concordant}
     \end{subfigure}
     \begin{subfigure}{.49\linewidth}
         \centering
         \includegraphics[width=.99\linewidth]{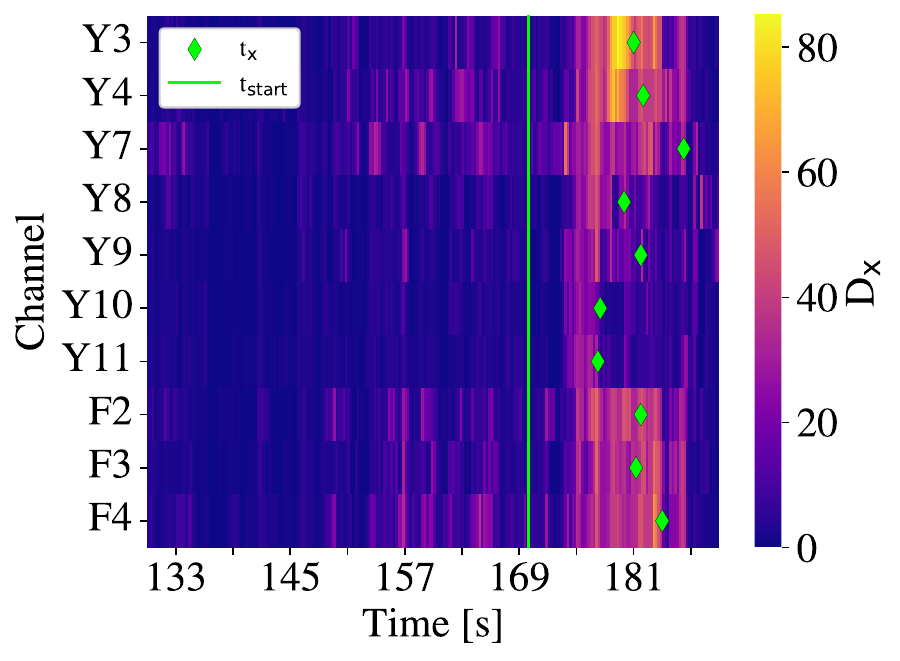}
         \caption{Desynchronization.}
         \label{fig:connection_concordant}
     \end{subfigure}
     \caption{The \gls{di} algorithm (patient $1$, $\eta=40\%$).}
     \label{fig:di_concordant}
\end{figure}

\begin{figure}[h!]
    \centering
    \begin{subfigure}{.49\linewidth}
        \centering
        \includegraphics[width=\linewidth]{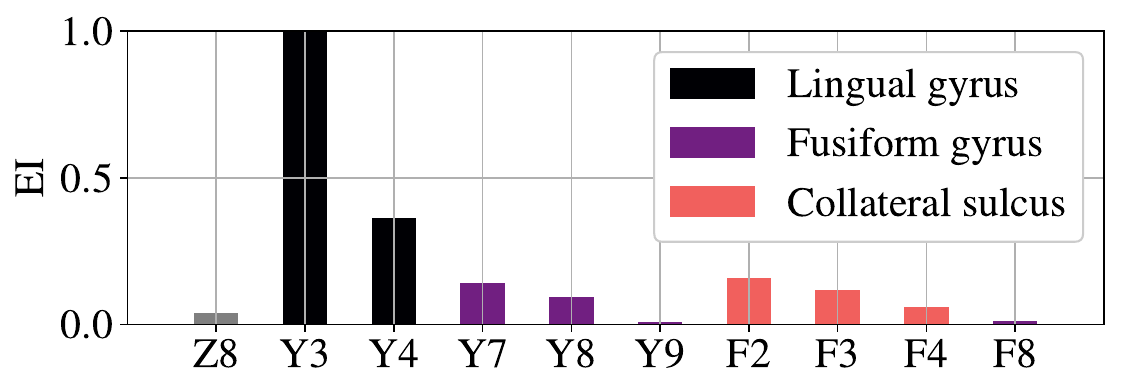}
        \caption{EI.}
        \label{fig:energy_index_concordant}
    \end{subfigure}
    \begin{subfigure}{.49\linewidth}
        \centering
        \includegraphics[width=\linewidth]{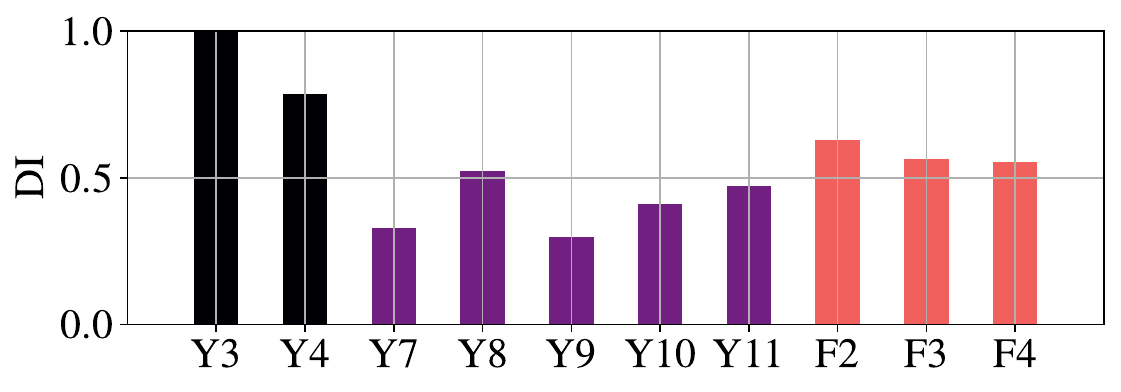}
        \caption{DI.}
        \label{fig:connection_index_concordant}
    \end{subfigure}
    \caption{EI and DI values for patient $1$.}
    \label{fig:ei_di_concordant}
\end{figure}

In Fig.~\ref{fig:di_concordant}, we investigate the results of the \gls{di} algorithm in the same scenario.
To ensure a fair comparison of the operative behavior of the algorithms, we set $\eta=40\%$, so that \gls{di} identifies the same number of epileptogenic channels as reported by \gls{ei}.
This allows us to examine how the algorithms select channels, regardless of differences in output size.
Finally, Fig.~\ref{fig:ei_di_concordant} reports the epileptogenicity levels assigned to each channel by the two algorithms, as well as the brain structure with which each channel is associated.
In both cases, the indices are normalized within the $[0, 1]$ range, such that values close to $1$ indicate a greater influence on the generation of epileptic discharge.

Overall, the results are largely concordant: both algorithms locate the seizure triggers within the lingual gyrus, the collateral sulcus and the fusiform gyrus.
Notably, all these regions were clinically identified as part of the \gls{ez} and were thermocoagulated at the end of \gls{seeg} monitoring.
In particular, the target patient underwent a cortectomy of the right lingual and fusiform gyri~\footnote{We specify that these clinical details are not derived from Tab.~I, which only reports the coarse EZ localization, but are based on the surgical reports associated with the patient.}, which results in a favorable seizure outcome (Engel II).

In detail, Figs.~\ref{fig:energy_concordant} and~\ref{fig:energy_index_concordant} show that \gls{ei} seems to identify channel $Y3$ as the primary epileptogenic focus.
Hence, $Y3$ visually exhibits the most abnormal energy variation after seizure onset ($t > 170.0$ s), and its $EI$ value exceeds that of all other channels by more than $50\%$.
Figs.~\ref{fig:connection_concordant} and~\ref{fig:connection_index_concordant} indicate that \gls{di} produces more diffuse outputs, with significant abnormal connectivity patterns distributed across channels $Y7$–$Y11$ and $F2$–$F4$, all of which report significant \gls{di} values.
However, the channel that exhibits the most prominent variation is still $Y3$, confirming its central role in originating epileptic discharges.

\subsection{Discordant Example}
\label{sub:discordant}

\begin{figure}[h!]
     \centering
     \begin{subfigure}{.49\linewidth}
         \centering
         \includegraphics[width=.99\linewidth]{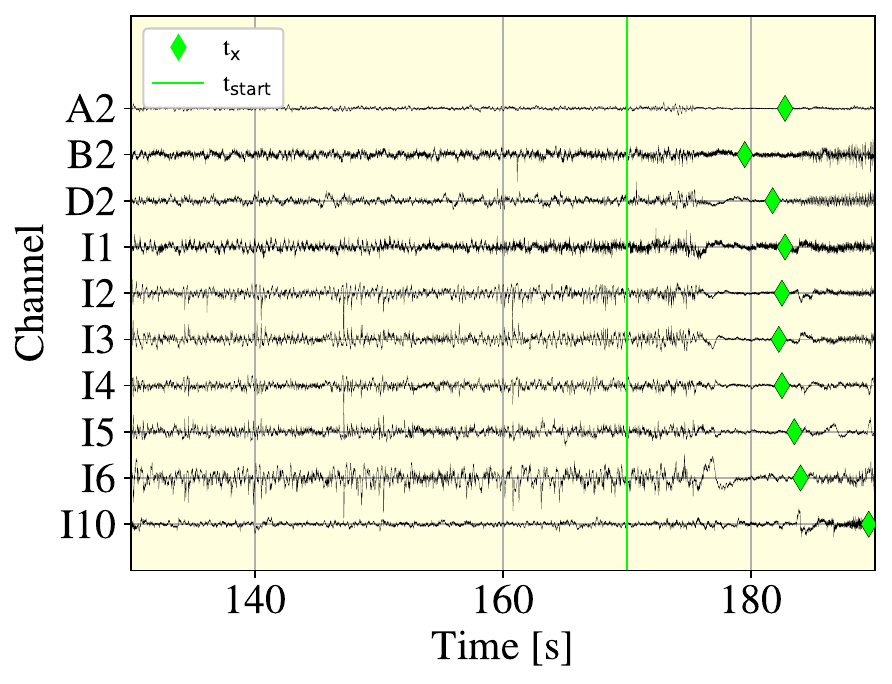}
         \caption{Signal.}
         \label{fig:energy_signal_discordant}
     \end{subfigure}
     \begin{subfigure}{.49\linewidth}
         \centering
         \includegraphics[width=.99\linewidth]{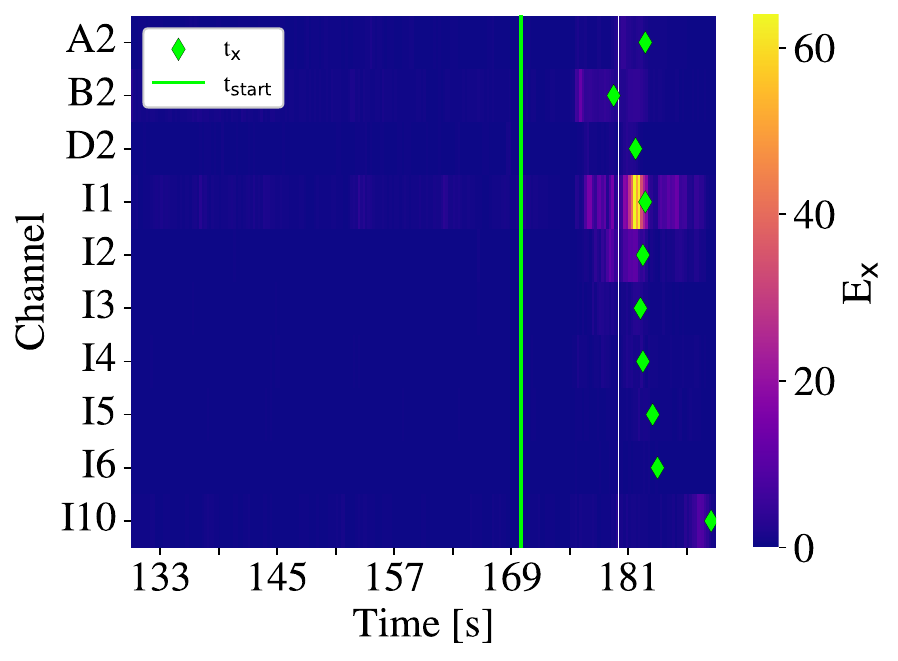}
         \caption{Energy ratio.}
     \label{fig:energy_discordant}
     \end{subfigure}
     \caption{The \gls{ei} algorithm (patient $2$, $\eta=6.5\%$).}
     \label{fig:ei_discordant}
\end{figure}

\begin{figure}[h!]
     \centering
     \begin{subfigure}{.49\linewidth}
         \centering
         \includegraphics[width=.99\linewidth]{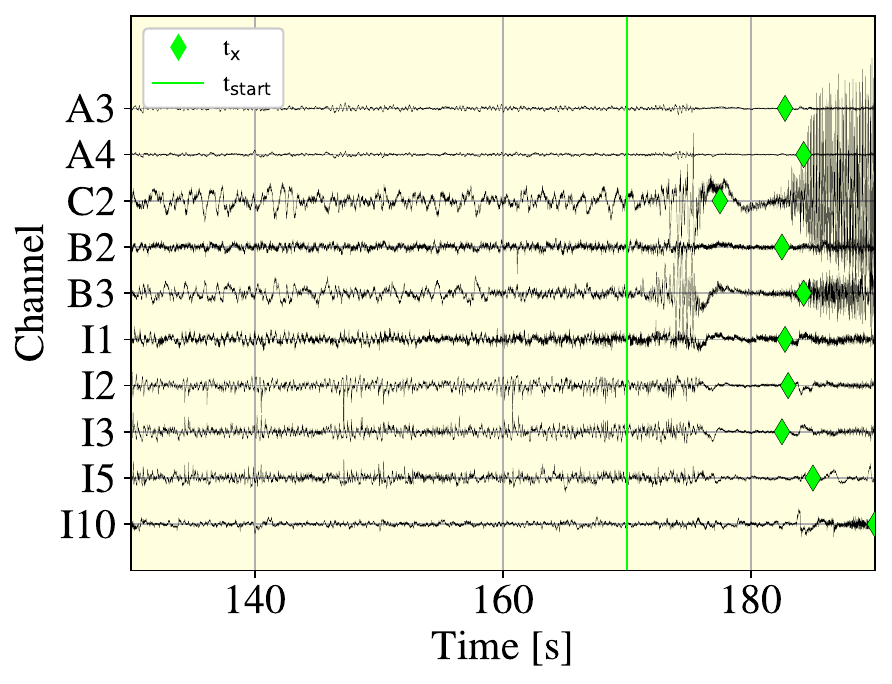}
         \caption{Signal.}
         \label{fig:connection_signal_disconcordant}
     \end{subfigure}
     \begin{subfigure}{.49\linewidth}
         \centering
         \includegraphics[width=.99\linewidth]{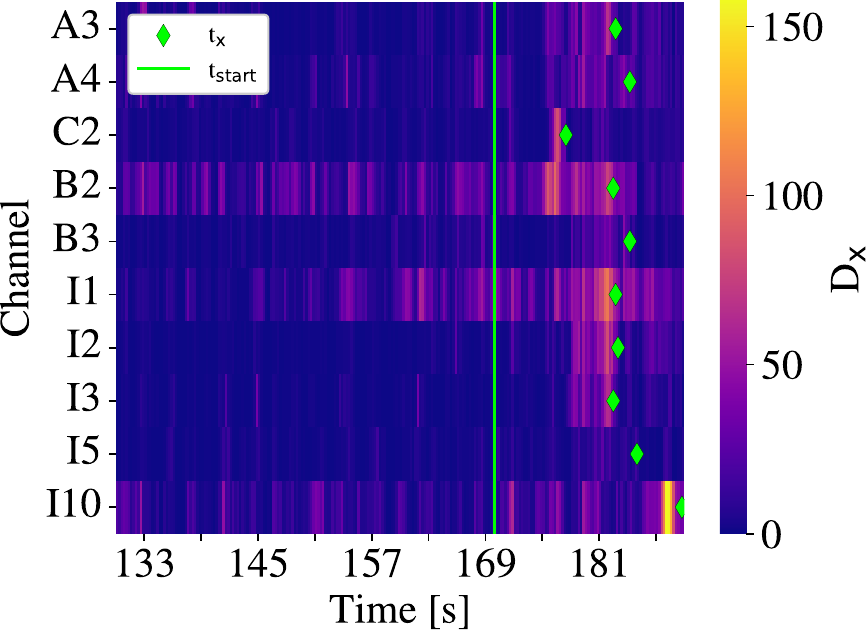}
         \caption{Desynchronization.}
         \label{fig:connection_disconcordant}
     \end{subfigure}
     \caption{The \gls{di} algorithm (patient $2$, $\eta=11\%$).}
     \label{fig:di_discordant}
\end{figure}

\begin{figure}[h!]
    \centering
    \begin{subfigure}{.49\linewidth}
        \centering
        \includegraphics[width=\linewidth]{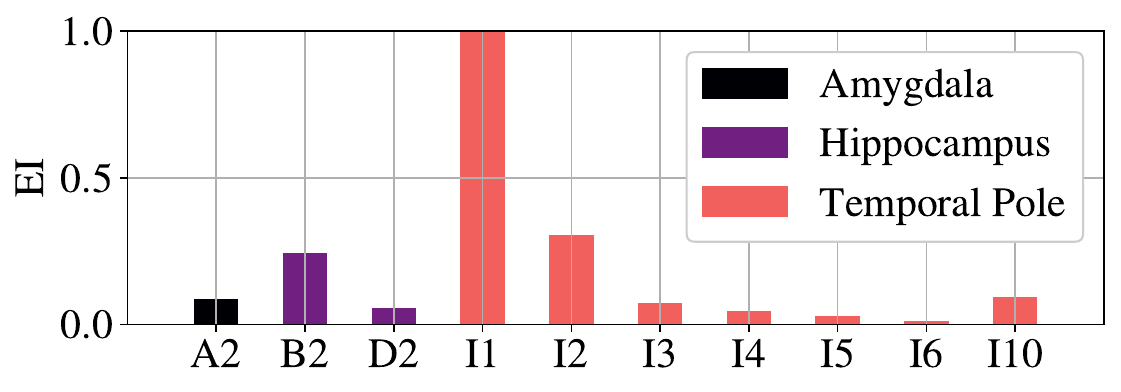}
        \caption{EI.}
        \label{fig:ei_output_discordant}
    \end{subfigure}
    \begin{subfigure}{.49\linewidth}
        \centering
        \includegraphics[width=\linewidth]{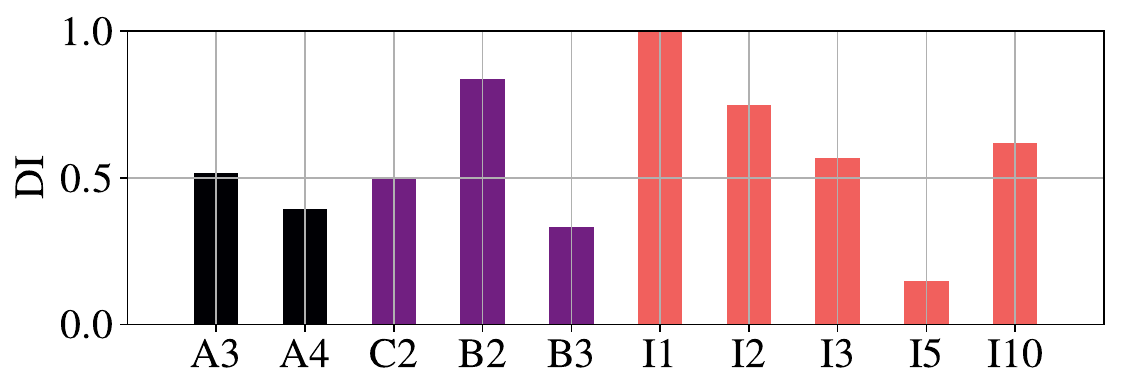}
        \caption{DI.}
        \label{fig:di_output_discordant}
    \end{subfigure}
    \caption{EI and DI values for patient $2$.}
    \label{fig:ei_di_discordant}
\end{figure}

In the following, we consider patient $2$, whose \gls{seeg} implant (shown in Fig.~\ref{fig:implant_2}) explores the left temporo-parietal and frontal areas and includes $163$ recording sites ($13$ electrodes).
Figs.~\ref{fig:ei_discordant} and~\ref{fig:di_discordant} report the channels considered the most epileptogenic according to the two algorithms together with the associated energy ratio and desynchronization values.
As before, the detection threshold $\eta$ was tuned to obtain the same number of outputs in the two cases, thus enabling a fair performance comparison.
Finally, Fig.~\ref{fig:ei_di_discordant} compares the epileptogenicity levels assigned by the two algorithms and the channel locations within the \gls{seeg} implant.

Figs.~\ref{fig:energy_discordant} and~\ref{fig:ei_output_discordant} indicate that \gls{ei} identifies channels $I1$–$I2$, exploring the mesial temporal pole, as the origin of epileptic discharges, being the only channels with prominent epileptogenicity values.
In contrast, Figs.~\ref{fig:connection_disconcordant} and~\ref{fig:di_output_discordant} show that \gls{di} produces more diffuse outputs, assigning elevated epileptogenicity values to multiple channels within the limbic system.
In particular, \gls{di} identifies epileptogenic phenomena in channels $C2$ and $B3$, which both explore the hippocampus and, critically, were not labeled as part of the \gls{ez} by the \gls{ei} algorithm.

As indicated by Tab.~\ref{tab:clinical}, the target patient (patient $2$) was diagnosed with hippocampal sclerosis and achieved seizure freedom only after the removal of both amygdala (channels $A1$-$A3$) and the hippocampus (channels $B1$-$B3$ and $C1$-$C3$).
This outcome suggests that, in this specific scenario, relying solely on the energy ratio could increase the risk of false negative errors. 
Therefore, integrating both energy- and connectivity-based biomarkers is essential to improve accuracy in the \gls{ez} definition and, consequently, the likelihood of seizure freedom.

\subsection{Aggregate Results}
\label{sub:aggregate}

\begin{figure}[h!]
     \centering
     \begin{subfigure}{.49\linewidth}
        \centering
        \includegraphics[width=\linewidth]{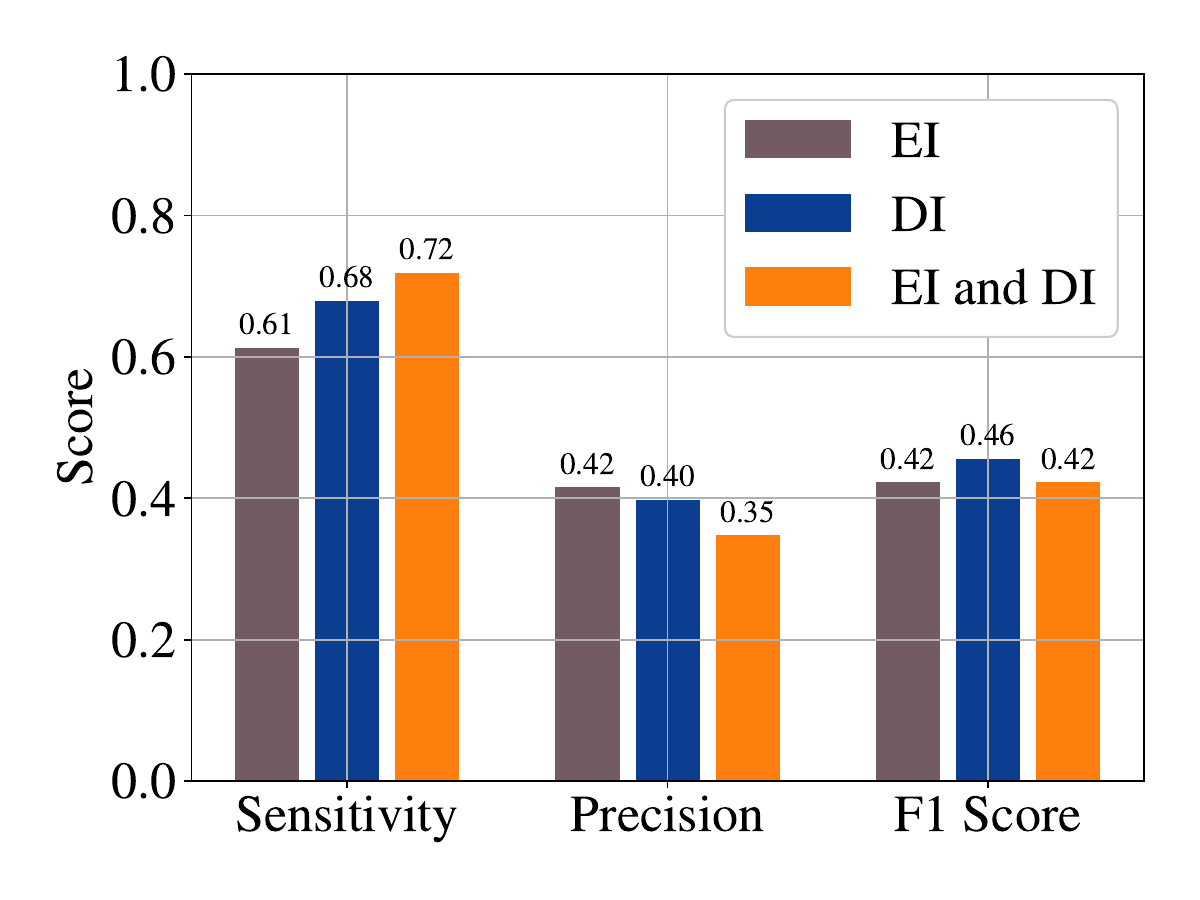}
        \caption{$\eta=5\%$.}
        \label{fig:score_epileptogenic_zone_10}
     \end{subfigure}
     \begin{subfigure}{.49\linewidth}
        \centering
        \includegraphics[width=\linewidth]{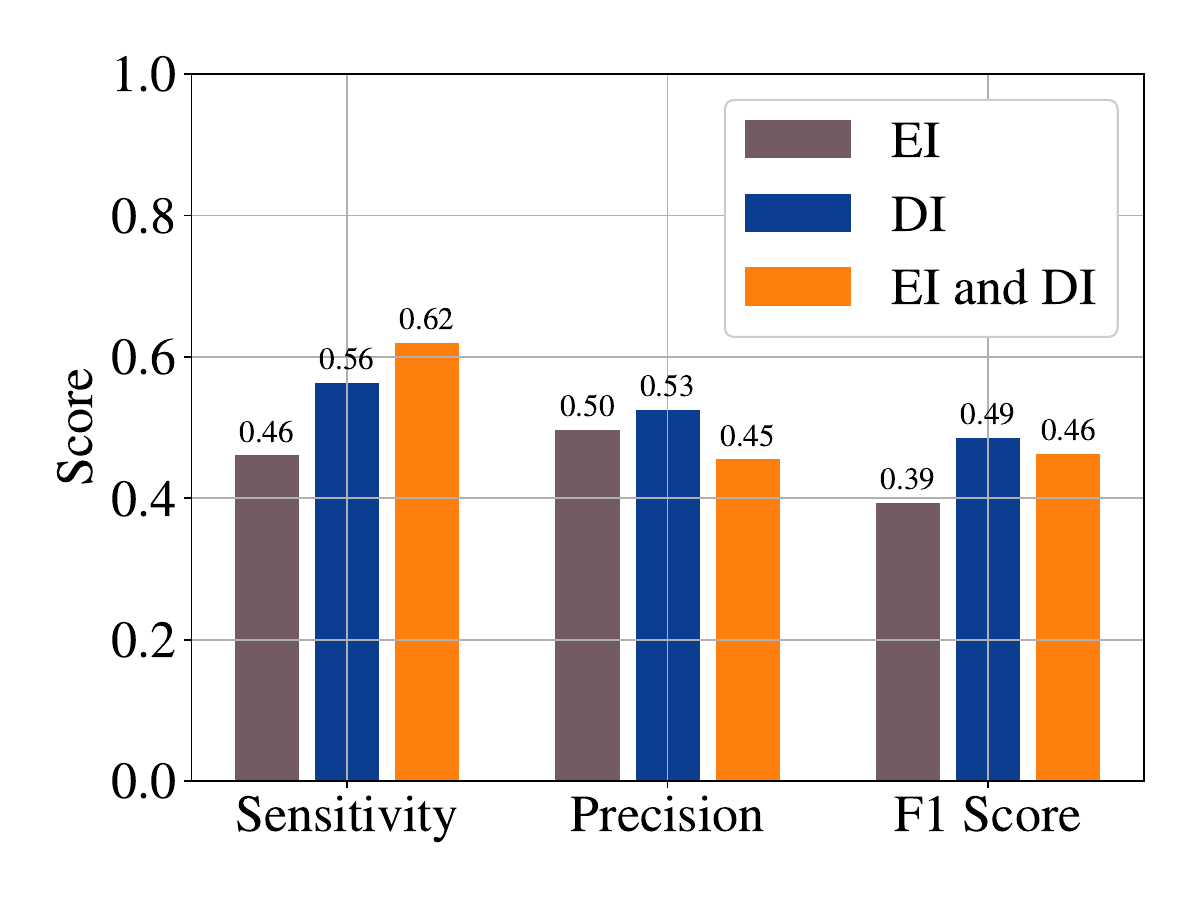}
        \caption{$\eta=10\%$.}
        \label{fig:score_epileptogenic_zone_20}
     \end{subfigure}
     \caption{Performance for different detection thresholds.}
     \label{fig:score_epileptogenic_zone}
\end{figure}

In the following, we analyze the performance of the \gls{ei} and \gls{di} algorithms across the entire patient population.
In particular, we consider each algorithm as a standalone tool, as well as their combination (EI and DI), in which a channel is considered part of the \gls{ez} if it is identified as positive by either \gls{ei} or \gls{di}.
Fig.~\ref{fig:score_epileptogenic_zone} reports the detection performance of the algorithms while setting $\eta=5\%$ and $\eta=10\%$ as detection threshold. 
In the first case, \gls{di} increases the sensitivity by $11.5\%$ compared to \gls{ei} ($0.68$ vs $0.61$), while the algorithm combination achieves a score of $0.72$.
This indicates that two algorithms produce different results and the channels with higher \gls{ei} values do not necessarily correspond to those with higher \gls{di} values.
Focusing on the F1 score, the \gls{di} algorithm outperforms both the benchmark and the joint configuration, achieving $\text{F1 score}=0.46$.

To improve precision, we can increase the threshold $\eta$, thus reducing the number of channels classified as part of the \gls{ez}. 
Fig.~\ref{fig:score_epileptogenic_zone_20} reports the results for $\eta=10\%$, showing that the \gls{ei} and the \gls{di} algorithms yield precisions of $0.50$ and $0.53$, respectively.
Taking into account the F1 score, \gls{di} improves performance compared to the previous scenario ($\text{F1 score}=0.49$) and still represents the best solution.
On the other hand, \gls{ei} and \gls{di} lead to $\text{F1 score}=0.39$ and $\text{F1 score}=0.46$, respectively.
We observe that the relatively low F1 scores are mainly due to the need to adjust the detection threshold $\eta$ individually for each patient, based on the extent of their \gls{seeg} implantation.

\begin{figure}[h!]
     \centering
     \begin{subfigure}{.49\linewidth}
        \centering
        \includegraphics[width=\linewidth]{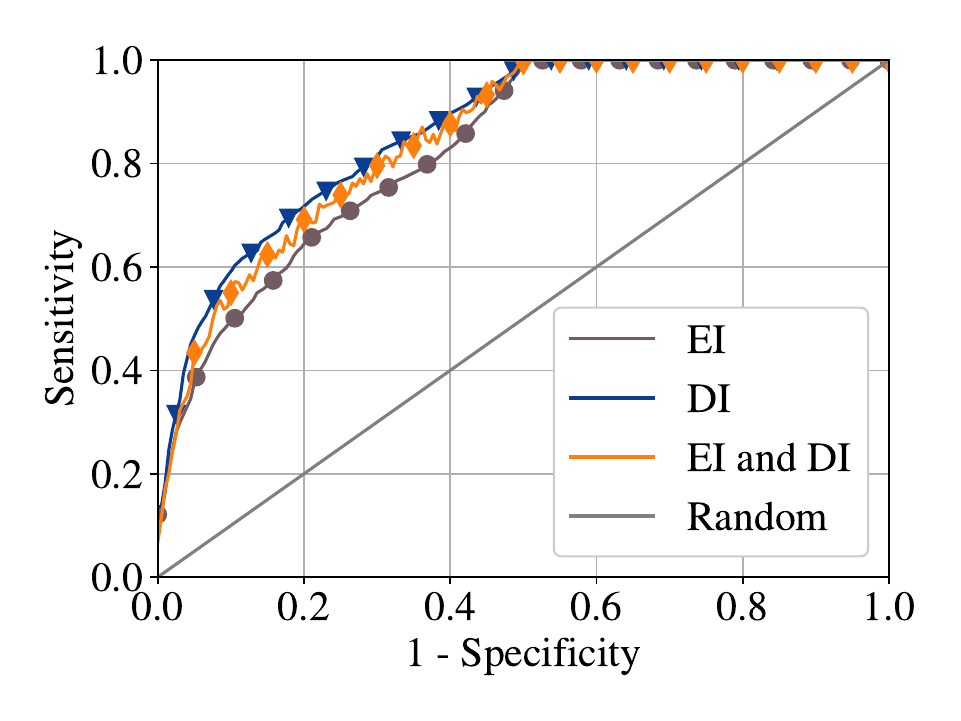}
        \caption{Receiver operating characteristic (ROC) curve.}
        \label{fig:roc_auc}
     \end{subfigure}
     \begin{subfigure}{.49\linewidth}
        \centering
        \includegraphics[width=\linewidth]{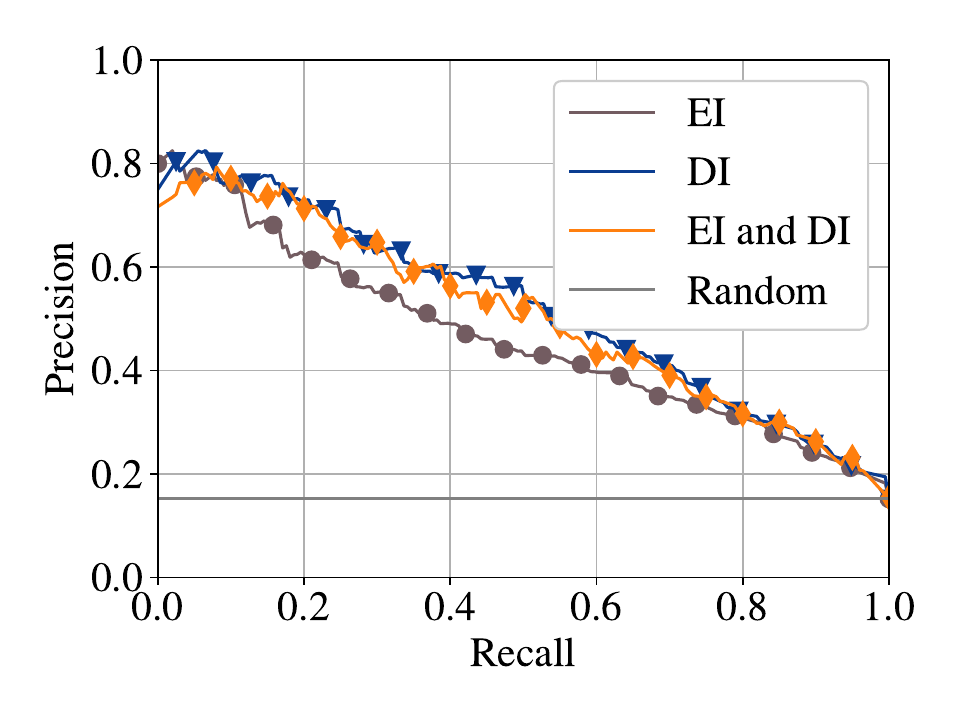}
        \caption{Precision-recall (PR) curve.}
        \label{fig:pr_auc}
     \end{subfigure}
     \caption{Trade-off between false positives and false negatives.}
     \label{fig:auc_epileptogenic_zone}
\end{figure}

To analyze the trade-off between sensitivity and specificity, we consider the average \gls{auc} across the clinical dataset.
Fig.~\ref{fig:roc_auc} denotes how the algorithm combination leads to the highest detection performance, with a final score of $\text{ROC-AUC}=0.88$, while the standalone algorithms yield $\text{ROC-AUC}=0.86$ for \gls{di} and $\text{ROC-AUC}=0.83$ for \gls{ei}. 
In particular, combining \gls{ei} and \gls{di} improves the \gls{auc} score by $5.2\%$ with respect to the \gls{ei} algorithm ($95\%$ confidence interval: $2.3-8.0\%$), enabling a more precise characterization of the ictal transition.
The benefit of integrating the two approaches is confirmed from the comparison of precision and the recall scores, as shown in Fig.~\ref{fig:pr_auc}.
The combined algorithm achieves $\text{PR-AUC}=0.53$, which is significantly higher than both \gls{ei} (PR-AUC = $0.43$) and \gls{di} (PR-AUC = $0.49$)
\footnote{We observe that, in the figure, the mean ROC and PR curves of the combined algorithm lie below those of the \gls{di} algorithm.
However, the expectations of the ROC-AUC and PR-AUC of the algorithm combination are higher than that of standalone \gls{di}.
This discrepancy arises because the mean curves were obtained by interpolating the values of the individual curves for each patient, which can distort the visual representation, whereas the mean ROC-AUC and PR-AUC values provide a more accurate measure of overall performance.}.

\begin{figure}[h!]
     \centering
     \begin{subfigure}{.49\linewidth}
        \centering
        \includegraphics[width=\linewidth]{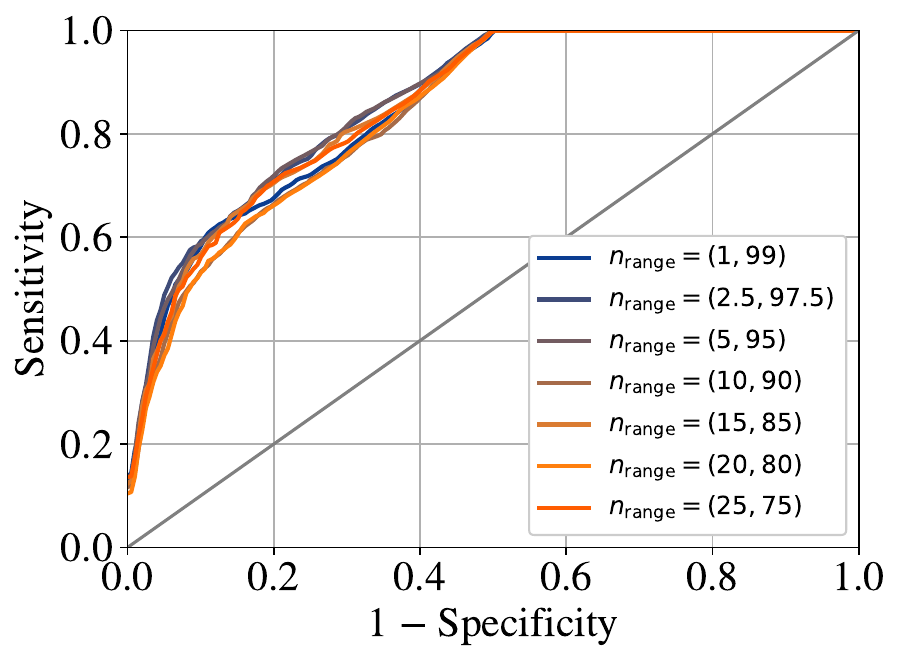}
        \caption{Receiver operating characteristic (ROC) curve.}
        \label{fig:desync_roc_auc}
     \end{subfigure}
     \begin{subfigure}{.49\linewidth}
        \centering
        \includegraphics[width=\linewidth]{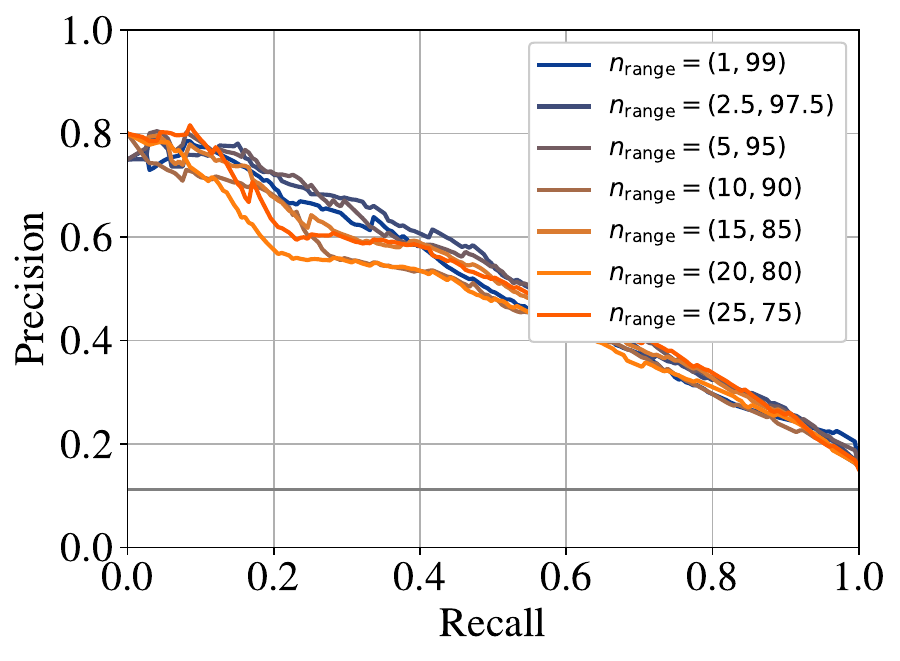}
        \caption{Precision-recall (PR) curve.}
        \label{fig:desync_pr_auc}
     \end{subfigure}
     \caption{Sensitivity analysis for the Desynchronization Index (DI) algorithm.}
     \label{fig:desync_sensitivity}
\end{figure}

Finally, we conducted a sensitivity analysis on the thresholding method used to distinguish spurious from authentic connectivity phenomena in Sec.~\ref{sec:desync}. Specifically, we computed \gls{di} performance in terms of \gls{auc} and \gls{pr-auc} by varying the $(n_{\text{low}}, n_{\text{high}})$ thresholds across the set $\{(1,99),$ $(2.5, 97.5),$ $(5, 95),$ $(10, 90),$ $(15, 85),$ $(20, 80),$ $(25, 75)\}$. 
As shown in Fig.~\ref{fig:desync_sensitivity}, detection performance remains consistent across different thresholds, ranging from a minimum $\text{\gls{auc}}=0.84$ for $(n_{\text{low}}, n_{\text{high}})=(20,80)$ to a maximum $\text{\gls{auc}}=0.87$ for $(n_{\text{low}}, n_{\text{high}})=(2.5,97.5)$.
The corresponding \gls{pr-auc} values are $\text{\gls{pr-auc}}=0.47$ and $\text{\gls{pr-auc}}=0.52$, respectively.

We observe that the best results are obtained by discarding connectivity values that are not statistically significant, using a two-sided test against a null distribution computed from the overall connectivity distribution in the \gls{seeg} time-varying network.
This demonstrates that our approach implicitly reduces the effect of volume conduction, effectively removing spurious connectivity from the analysis.
A summary of the specific performance of each algorithm, including \gls{ei} and the combined configuration, is reported in Tab.~\ref{tab:performance}. Notably, the table reports the sensitivity and specificity values corresponding to the best F1 score, obtained while exploring all detection thresholds $\eta \in [0,1]$.

\begin{table*}[t!]
\footnotesize
\centering
\caption{Performance metrics for different algorithms.}
\label{tab:performance}
\begin{tabular}{cccccccc} 
\hline
\textbf{Algorithm} & $\bm{n_{\text{low}}, n_{\text{high}}}$ & \textbf{ROC-AUC} & \textbf{PR-AUC} & \textbf{F1-Score} & \textbf{Sensitivity} & \textbf{Specificity} \\ 
\hline
\multirow{7}{*}{DI} 
   & $(1, 99)$ & 0.85 & 0.50 & 0.55 & 0.58 & 0.58 \\
   & $(2.5, 97.5)$ & 0.87 & 0.52 & 0.57 & 0.59 & $\bm{0.61}$ \\
   & $(5, 95)$ & 0.86 & 0.49 & 0.57 & 0.61 & 0.58 \\
   & $(10, 90)$ & 0.84 & 0.48 & 0.53 & 0.61 & 0.52 \\
   & $(15, 85)$ & 0.86 & 0.51 & 0.55 & 0.58 & 0.57 \\
   & $(20, 80)$ & 0.84 & 0.47 & 0.51 & $\bm{0.65}$ & 0.49 \\
   & $(25, 75)$ & 0.86 & 0.50 & 0.54 & $\bm{0.65}$ & 0.53 \\ 
EI & \verb|\| & 0.83 & 0.43 & 0.50 & 0.59 & 0.51 \\ 
EI and DI & \verb|\| & $\bm{0.88}$ & $\bm{0.53}$ & $\bm{0.58}$ & 0.62 & 0.50 \\ 
\hline
\end{tabular}
\end{table*}

\section{Clinical Discussion}
\label{sec:discuss}

The purpose of this study is to validate the hypothesis that the generation of epileptic discharges is associated with a \emph{desynchronization} between the \gls{ez} and the rest of the brain.
This hypothesis is supported by prior studies~\cite{bartolomei2001neural}, but to our knowledge, has not yet been explicitly applied to identify epileptogenic channels in \gls{seeg} signals, as implemented by the \gls{di} algorithm.
The results of the work support our initial hypothesis: \gls{di} outperforms state-of-the-art approaches in the localization of the \gls{ez}, reinforcing the use of channel desynchronization as a promising biomarker for the analysis of \gls{seeg} data.

As \gls{pte} estimates the synchrony between neural signals based on the distribution of the instantaneous phase, the \gls{di} algorithm identifies as epileptogenic the channels that exhibit a static phase distribution over time and thus are not influenced by other channels in the \gls{seeg} implant.
From an analytical perspective, our framework assumes that epileptogenic sites exhibit independent behavior that cannot be suppressed by other cortical structures.
In this vision, during the interictal phase, the regions surrounding the \gls{ez} exert an inhibitory effect, preventing the generation of seizures.
Hence, seizure onset occurs when the exchange of information between the epileptogenic and physiological regions is disrupted, consistent with the findings of recent studies~\cite{jiang2022interictal, awad016}.

In a clinical context, \gls{di} is not intended to directly plan the resection area for epilepsy surgery, but rather to support neurophysiologists in the \gls{ez} definition.
In particular, the combined use of \gls{di} and \gls{ei} can be used to build a first sketch of \gls{ez}, allowing neurophysiologists to look at a limited number of channels rather than the entire \gls{seeg} network.
This could be beneficial in the case of extensive \gls{seeg} implants, where understanding the dynamics of seizures is extremely difficult and can lead to fatal human errors. 
At the same time, the clinical judgment of neurophysiologists remains an essential element in discerning false and true positive, counterbalancing the lower precision of the proposed framework.

We observe that the \gls{di} algorithm may implicitly reflect the presence of ictal activity in cortical regions that are not directly sampled by the \gls{seeg} electrodes.
Since we consider desynchronization as the main biomarker of epileptogenicity, the mere presence of epileptiform waveforms, such as fast oscillations or spikes, diffusely across most \gls{seeg} channels is not enough to consider a channel as pathological. 
In other words, epileptiform activities that do not coincide with a concurrent desynchronization may reflect physiological brain processes or seizure propagation phenomena rather than the true \gls{ez}.

From a practical perspective, the presence of channels with high \gls{ei} values but no corresponding \gls{di} values may indicate that the seizure originates from a hidden structure.
Therefore, integrating the \gls{di} and \gls{ei} algorithms could distinguish true ictal patterns from the propagation of epileptic discharges, helping neurophysiologists determine whether the epileptogenic network is fully or only partially explored.
As a long term objective, the proposed framework could thus be expanded to assess whether the recorded \gls{seeg} data provide sufficient information to ensure a high likelihood of successful epilepsy surgery.

We note that our pipeline does not include any pre-filtering stage for the \gls{seeg} data, making connectivity information be extracted directly from broadband signals.
Consequently, unlike most state-of-the-art approaches, the \gls{di} algorithm estimates the connectivity of each \gls{seeg} channel considering all its oscillatory components.
This enables the algorithm to be implemented in a patient-agnostic manner, without requiring assumptions about the specific waveforms characterizing ictal phenomena, which, notably, may vary from case to case, depending on the location of the \gls{ez}~\cite{8632671}.
On the other hand, \gls{di} can also be tuned to specific frequency ranges, which may further enhance detection accuracy.

Notably, this manuscript considers \gls{rftc} channels as the ground truth for the detection framework.
In the studied population, thermocoagulated channels accurately targeted epileptogenic tissue, since all \gls{rftc} procedures were performed with curative intent and the analysis was restricted to patients who achieved favorable seizures (Engel I–II).
Nevertheless, several important limitations should be acknowledged: \gls{rftc} does not provide histopathological validation; spatial sampling is limited by electrode placement; and the selection of \gls{rftc} targets is influenced by the need to preserve eloquent cortex, rather than being based solely on epileptogenicity.

In this regard, we note that other studies in this field have often considered the surgically resected area as the clinical ground truth.
However, such an approach may not provide an accurate delineation of the \gls{ez}, as surgical resections typically cover a region broader than the \gls{ez} alone.
In our case, selecting thermocoagulated contacts as the detection target is more consistent with the clinical objective of the \gls{seeg} evaluation, which is to minimize the extent of cortical tissue resected during surgery and reduce the risk of damaging eloquent structures.

Finally, we observe that the main limitation of this study is the use of a relatively small dataset from a single clinical center.
Although our primary objective was to propose a methodology for identifying desynchronization phenomena, extending the analysis to larger datasets will be essential to validate the performance of the \gls{di} algorithm and assess its feasibility in real-world scenarios.

\section{Conclusions}
\label{sec:concl}

\glsreset{di}
\glsreset{ei}
\glsreset{ez}
\glsreset{auc}

This work presented a framework for analyzing \gls{seeg} signals and, ultimately, improving the likelihood of seizure freedom after epilepsy surgery.
We designed a new algorithm, termed \gls{di}, which quantifies the epileptogenic level of \gls{seeg} channels by measuring the reduction in outward connectivity during the ictal transition.
Accordingly, the algorithm considers the ability of a cortical site to behave independently, and thus to \emph{desynchronize} from the brain network, as a key biomarker of the \gls{ez}.

We evaluated DI on a dataset of $20$ patients and compared its performance against the \gls{ei}, which represents the state of the art for identifying the \gls{ez}.
The results showed that \gls{di} leads to a higher \gls{auc} than \gls{ei} ($\text{AUC}=0.86$ vs $\text{AUC}=0.83$), and combining the two tools improves the AUC by $5.2\%$ with respect to \gls{ei} (CI: $2.3-8.0\%$).

Our results confirm that the desynchronization of \gls{seeg} channels represents a key biomarker to identify structures within the \gls{ez}. 
In particular, \gls{di} underscores signal modifications that are not visually evident and, when combined with other quantitative biomarkers, may substantially improve the success rate of epilepsy surgery in complex scenarios. 
In future work, we intend to clinically validate the \gls{di} algorithm on a larger dataset, possibly including data from different clinical research centers, and to evaluate the potential of the proposed computational framework for analyzing the inter-ictal period.

\bibliographystyle{unsrt}

\end{document}